\documentclass[twocolumn,twocolappendix]{aastex631}
\usepackage{xspace}
\usepackage{amsmath}
\usepackage[multiple]{footmisc}
\usepackage{threeparttable}
\usepackage{CJK}

\newcommand{\chandra}{\textit{Chandra}\xspace}

\newcommand{\ciao}{\textit{CIAO}\xspace}
\newcommand{\marx}{\textit{MARX}\xspace}
\newcommand{\saotrace}{\textit{SAOTrace}\xspace}
\newcommand{\sherpa}{\textit{Sherpa}\xspace}
\newcommand{\ObsID}{\texttt{ObsID}\xspace}

\definecolor{salmon}{rgb}{1.0, 0.55, 0.41}
\definecolor{ultrapink}{rgb}{1.0, 0.44, 1.0}

\shorttitle{G292's pulsar proper motion}
\shortauthors{Xi et al.}
\graphicspath{{./}{figures/}}

\begin{document}
\begin{CJK*}{UTF8}{gbsn}
\title{The Proper Motion of the Pulsar J1124--5916 in the Galactic Supernova Remnant G292.0+1.8}

\author[0000-0003-3350-1832]{Xi Long}
\affiliation{Center for Astrophysics $|$ Harvard \& Smithsonian \\
60 Garden St., MS-3 \\
Cambridge, MA 02138, USA}
\author[0000-0002-7507-8115]{Daniel J. Patnaude}
\affiliation{Center for Astrophysics $|$ Harvard \& Smithsonian \\
60 Garden St., MS-3 \\
Cambridge, MA 02138, USA}
\author[0000-0003-1415-5823]{Paul P. Plucinsky}
\affiliation{Center for Astrophysics $|$ Harvard \& Smithsonian \\
60 Garden St., MS-3 \\
Cambridge, MA 02138, USA}
\author[0000-0002-5115-1533]{Terrance J. Gaetz}
\affiliation{Center for Astrophysics $|$ Harvard \& Smithsonian \\
60 Garden St., MS-3 \\
Cambridge, MA 02138, USA}



\begin{abstract}

We present the first direct measurement of the proper motion of pulsar J1124--5916 in the young, oxygen-rich supernova remnant G292.0+1.8. Using deep {\it Chandra} ACIS-I observations from 2006 and 2016, we measure a positional change of 0$\farcs$21 $\pm$ $0\farcs05$ over the $\sim$ 10 year baseline, or $\sim$ $0 \farcs 02$ yr$^{-1}$. At a distance of 6.2 $\pm$ 0.9 kpc, this corresponds to a kick velocity in the plane of the sky of $\mathrm{612\pm 152\,km \, s^{-1}}$. We compare this direct measurement against the velocity inferred from estimates based on the center of mass of the ejecta. Additionally, we use this new proper motion measurement to compare the motion of the neutron star to the center of expansion of the optically emitting ejecta. We derive an age estimate for the supernova remnant of $\gtrsim$ 2000 years. The high measured kick velocity is in line with recent studies of high proper motion neutron stars in other Galactic supernova remnants, and consistent with a hydrodynamic origin to the neutron star kick.

\end{abstract}



\section{Introduction} \label{sec:intro}

Neutron stars (NSs) are formed during the core-collapse supernovae (CCSNe) of stars with zero age main sequence masses of $\gtrsim$ 8M$_{\sun}$. Observations of neutron stars and pulsars indicate high space velocities with $v_{\mathrm{3D}}$ $\approx$  300--400 km s$^{-1}$ \citep{Hobbs2005,igoshev2020I}. There are currently believed to be two possible mechanisms which can accelerate neutron stars to these high velocities at birth: anisotropic neutrino emission \citep[e.g.,][]{Woosley1987,Socrates2005,Fryer2006}, and hydrodynamic kicks \citep[e.g.,][]{Janka1994,Burrows1996}. 

In the case of anisotropic neutrino emission, an anisotropy of only $\sim$1\% would be sufficient to impart a kick velocity of several hundred km s$^{-1}$ \citep[e.g.,][]{Socrates2005}, though \citet{Wongwathanarat2013} argue that an anisotropy of this size is difficult to achieve without invoking strong assumptions about the proto-neutron star. In particular, exotic theories concerning neutrino interactions, strong magnetic fields ($B > 10^{15}$ G) in the proto-neutron star, or turbulence in the neutrinosphere of the neutron star are required \citep{Wongwathanarat2010}. In contrast, recent two- and three-dimensional models for neutrino-driven core-collapse supernovae suggest that NS kicks of $\lesssim$ 1500 km s$^{-1}$ are achievable solely through bulk hydrodynamic kicks \citep[e.g.,][]{Nakamura2019}. By way of momentum conservation, the neutron star is imparted a momentum in the direction opposite to that of asymmetrically ejected stellar debris. On short timescales ($t$ $\sim$ 1 second) both hydrodynamic and gravitational forces work to impart momentum to the neutron star \citep{Wongwathanarat2010,Wongwathanarat2013}, while on longer timescales, additional momentum is imparted by gravity from high velocity ejecta \citep[the gravitational ``tugboat" effect;][]{Wongwathanarat2013}.

Recent studies by \citet{HollandAshford2017} and \citet{Katsuda2018} investigated, by independent means, the relationship between neutron star kick velocity and direction, and the location of iron-group and intermediate mass elements in the shocked supernova ejecta of several Galactic supernova remnants (SNRs). In both instances, the authors concluded that the measured or inferred kick velocities of the neutron stars in their sample were brought about by asymmetric explosions. In both papers, neutron star proper motions were incorporated either from a direct measurement of the proper motion or by measuring the offset of the neutron star from the geometric center of expansion (center of mass). However, in two cases, Cas A and G292.0+1.8, the authors note that they compute the proper motion based off of the offset of the neutron star from the optically determined center of expansion \citet{HollandAshford2017}. For G292.0+1.8, both papers assume a kick velocity of $\sim$ 450 km s$^{-1}$ \citep{Winkler2009}. 

G292.0+1.8 (hereafter G292) is a young ($t_{\mathrm{SNR}}$ $\sim$ 3000 year old, \citet{Winkler2009}) oxygen-rich SNR. The remnant is bright in X-ray emission from shocked ejecta and exhibits a prominent bar of shocked circumstellar material across the middle. A pulsar (J1124--5916) and bright pulsar wind nebula are associated with this remnant. The SNR is located at a distance of $\sim$ 6 kpc, and has an angular diameter of $\sim$ 8$\arcmin$ \citep{Gaensler2003}. The SNR is thought to be the remnant of a IIL/IIb supernova \citep{Chevalier2005}, with initial progenitor mass estimates of 13--30 M$_{\sun}$ \citep{Bhalerao2019}. X-ray studies of shocked circumstellar material suggest it is still expanding into the red supergiant wind \citep{Lee2010}. The mass loss rate during the red supergiant phase is estimated to be 2--5$\times$10$^{-5}$ M$_{\sun}$ yr$^{-1}$, with 15--40 M$_{\sun}$ of circumstellar material shock heated to X-ray temperatures \citep{Lee2010}. A recent analysis by \citet{Jacovich2021} placed tighter constraints on the progenitor properties, suggesting an initial mass of $\lesssim$ 20 M$_{\sun}$, though the estimated mass loss rate from \citet{Lee2010} is broadly consistent with values from \citet{Jacovich2021} for a 20 M$_{\sun}$ progenitor. \citet{Bhalerao2019} performed a fairly complete census of the X-ray emitting material, and derived a total ejecta mass of $\sim$ 6 M$_{\sun}$. 

Here we report on the first direct measurement of the proper motion of J1124--5916 in G292. Combining deep multi-epoch {\it Chandra} observations registered against the {\it Gaia} Data Release 3, we are able to correct the astrometry to an accuracy of 50 milliarcseconds (mas). In Sections~\ref{sec:data} and~\ref{sec:registration}, we present the {\it Chandra} observations and our registration technique; in Section~\ref{sec:propermotion} we present the proper motion measurement. In Section~\ref{sec:discussion}, we discuss our results, and attempt to place the measured proper motion of J1124--5916 in the context of other neutron stars. 

\section{Data and Reduction} \label{sec:data}

G292 has been observed several times over the course of the {\it Chandra} mission. Here we make use of two large programs from 2006 and 2016. Our goal is to match faint point sources detected in the {\it Chandra} observations against known sources from the Gaia 3rd Data Release (DR3). In order to
minimize any cumulative error which we might incur when adding several shorter {\it Chandra} observations performed at different roll angles and exposure times, we limited ourselves to the longest observations from each program, all performed at a similar roll angle, in order to minimize any systematic error. The observations are listed in Table~\ref{tab:obslist}, where we also list relevant observation information, and the $\Delta t$ between the individual exposure and the Gaia DR3 reference epoch of $2016.0$ (denoted as $\Delta t_{\mathrm{2016.0}}$).

\subsection{Relevant Software and Archives} \label{sec:tools}

Using the \ciao tool \texttt{chandra\_repro}\footnote{https://cxc.harvard.edu/ciao/ahelp/chandra\_repro.html} and CALDB 4.9.5, we reprocessed each observation to generate new L2 event lists. \marx\footnote{https://space.mit.edu/cxc/marx/} and \saotrace\footnote{https://cxc.cfa.harvard.edu/cal/Hrma/SAOTrace.html} were used for simulating the point spread function (PSF) of point sources. The positions of the point sources, as determined by the \ciao tool \texttt{wavedetect} were fitted using the PSF image with \sherpa, following the CIAO thread that discusses how to account for PSF effects in 2D image fitting\footnote{https://cxc.harvard.edu/sherpa/threads/2dpsf/}. To correct for time dependent changes to the quantum efficiency and exposure time differences among observations event-by-event, we used the \ciao tool \texttt{eff2evt}, by setting the option \texttt{detsubsysmod} to the start time of the observation, and the start time of the reference observation, \ObsID 19892. Finally, we used the GAIA Archive\footnote{https://gea.esac.esa.int/archive/} to search for optical counterparts to detected X-ray sources. The matched optical point sources are used as a reference frame to register all the observations.
\begin{table}[htb]
\caption{Observation List}
\begin{center}
\label{tab:obslist}
\begin{tabular}{ l l c c c}
\hline\hline
ObsID & Start Date & Exposure & Roll Angle & $\Delta t_{2016.0} $\\ & & (ks) & & (yr)\\
\hline 
\phantom{0}6677 & 2006-10-16 & 159.13 & 140.19 & 9.21 \\
\phantom{0}6679 & 2006-10-03 & 153.95 & 156.69 & 9.24 \\
\phantom{0}8221 & 2006-10-20 & \phantom{0}64.96 & 140.19 & 9.20 \\
19892 & 2016-10-05 & \phantom{0}49.48 & 150.19 & -0.76 \\
19899 & 2016-10-18 & \phantom{0}42.57 & 144.19 & -0.80 \\
\hline
\end{tabular}
\end{center}
\end{table}

\subsection{Position of X-ray point sources} \label{sec:pointsource}
To determine the position of the X-ray sources, we use a PSF 
fitting method based on the CIAO thread on 2D image fitting described in the 
previous section.  We include more details specific to our analysis in the
discussion below. We first create an image of the source to be modelled 
from a $31\farcs5 \times 31\farcs5$ box region, which we refer to as the
``data image''.  We then use \saotrace to simulate the PSF  at the
off-axis position  appropriate for each point source of interest
with a power-law spectrum with an index of 2.94 consistent 
with the average spectral properties of the sources used for 
registration. We scale the source flux by a factor of five thousand to reduce the statistical noise. This simulation is used to create an image of a point source at this position in the focal plane.
We convolve this PSF model image with a Gaussian plus a constant. The convolution with a Gaussian smooths the statistical fluctuations in the
simulated data, ensures that the values are strictly positive, and allows for
interpolation between pixels to non-integer values. This smoothing is 
necessary to prevent the possible source positions from being quantized to 
the original grid of values determined by the detector coordinate system. The
constant term accounts for the background in the real data.  We refer to this
image as the ``model image''. We then fit the model image to the data image (both binned to 1/4 sky pixels).
 The center, amplitude, and sigma of the 
Gaussian function, and the constant value are free 
to vary in the fit. We employed the C-statistic \citep{Cash1979} to determine 
the best fit value of the parameters. The fitted center of the Gaussian 
function $(x,y)$ is the position of the point source in image coordinates. The coordinates are transformed to sky coordinates 
using the WCS tranformations.  The $1 \sigma$ confidence intervals are 
determined using the sherpa 
\texttt{conf}\footnote{https://cxc.harvard.edu/sherpa/ahelp/conf.html} 
routine, which finds the root of the C-stat function \textit{vs} $x$ or $y$: 
$c(x)-(c_{\mathit{min}} +1)=0$ or $c(y)-(c_{\mathit{min}} +1)=0$. The root of 
the C-stat function \textit{vs} $x$ or $y$ is determined using Muller's 
method. An approximate root $x_\mathit{k}$ is generated for each iteration 
$k$; the iteration is halted when $x_\mathit{k+1} - x_\mathit{k} < 0.01$; 
$x_\mathit{k+1}$ is then the root of $c(x) - (c_{\mathit{min}}+1) = 0$.
The sigma of the Gaussian must remain small in our fits in order to avoid broadening the model image significantly beyond the resolution of the Chandra
PSF. The fitted values of $\sigma$ were never larger than 0\farcs35 in our fits. The point sources were all at off-axis angles $\ga4\arcmin$. The Chandra 90\% encircled energy is $2\farcs0$ at an off-axis angle of 4\farcm0. The additional broadening due to our Gaussian smoothing is small or negligible compared to the scale of the Chandra PSF.

\begin{figure}
    \centering
    \includegraphics[width=0.45\textwidth]{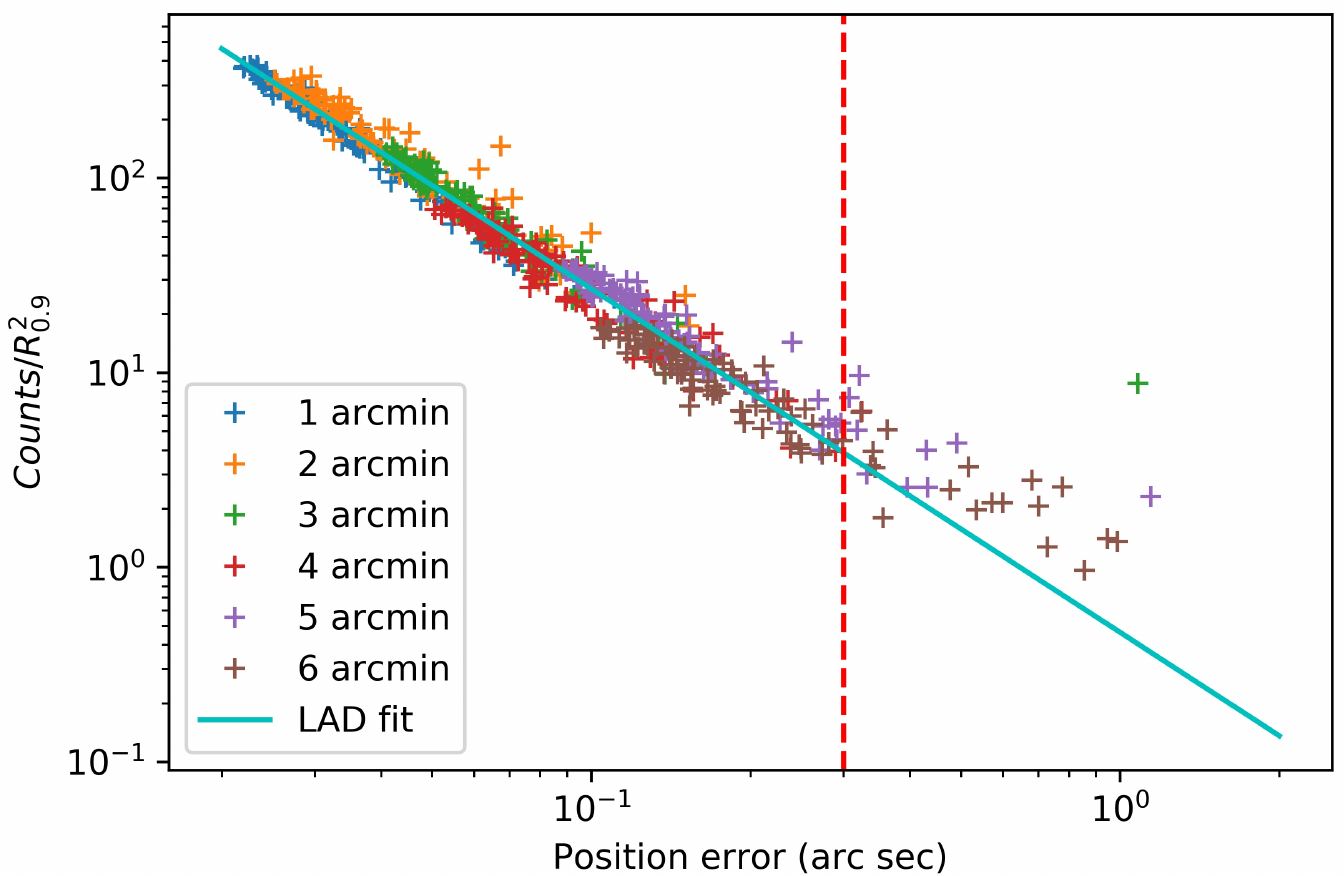}
    \caption{The ratio of counts to 90\% enclosed area of the simulated point sources versus the fitted positional error. The cyan line is the fitted power-law model using the least absolute deviation (LAD) method. The red dashed line indicates a positional error of $0\farcs3$. Those point sources with a positional error larger than $0\farcs3$ indicate that the relationship between positional error, surface brightness, and off-axis angle is not well described by a power-law model beyond $0\farcs3$. For high precision source registration, these sources would be excluded, as nonlinear effects appear to influence the measured positional error. \label{fig:positionerror}}
\end{figure}

The measured error in an X-ray source's position depends upon at least the number of counts in the source and the size and shape of the PSF, which depends primarily on the off-axis angle of the source and roll angle of the telescope. In Table~\ref{tab:counts}, we list the number of counts, off-axis angle and the positional errors for each of the X-ray sources.

\begin{table*}[htb]
\caption{Point source and the pulsar counts ($C$), off-axis angles ($\theta$) and the fitted positional errors $\sigma_{P}$ of the point sources.}
\begin{center}
\begin{tabular}{ l|c c c c c c c c c c c c c c c}
\hline\hline
& &6677 & & & 6679 & & &8221 & & &19892 & & &19899& \\
Source ID$^{\dagger}$     & $C$ & $\theta$ & $\sigma_{P}$ &  $C$ & $\theta$ & $\sigma_{P}$& $C$ & $\theta$ & $\sigma_{P}$& $C$ & $\theta$ & $\sigma_{P}$& $C$ & $\theta$ & $\sigma_{P}$\\
\hline
1 & 1263 & $7.0\arcmin$ & 0.04\arcsec & 1201 & $6.9\arcmin$ & 0.04\arcsec & 529 & $7.1\arcmin$ & 0.06\arcsec & 388 & $6.8\arcmin$ & 0.07\arcsec & 398 & $6.8\arcmin$& 0.06\arcsec\\
2 & 1485 & $8.6\arcmin$ & 0.05\arcsec & 1126 & $8.5\arcmin$ & 0.05\arcsec & 465 & $8.7\arcmin$ & 0.08\arcsec & 643 & $8.3\arcmin$ & 0.06\arcsec & 599 & $8.4\arcmin$& 0.09\arcsec\\
3 &   99 & $4.5\arcmin$ & 0.14\arcsec &   92 & $4.6\arcmin$ & 0.17\arcsec &  51 & $4.5\arcmin$ & 0.21\arcsec &  20 & $4.8\arcmin$ & 0.21\arcsec &  18 & $4.7\arcmin$& 0.27\arcsec\\
4 &  197 & $4.4\arcmin$ & 0.10\arcsec &  130 & $4.6\arcmin$ & 0.09\arcsec &  86 & $4.4\arcmin$ & 0.10\arcsec &  25 & $4.7\arcmin$ & 0.26\arcsec &  23 & $4.6\arcmin$& 0.19\arcsec\\
5 &  304 & $5.2\arcmin$ & 0.07\arcsec &  148 & $5.4\arcmin$ & 0.09\arcsec &  57 & $5.2\arcmin$ & 0.17\arcsec &  36 & $5.5\arcmin$ & 0.29\arcsec &  23 & $5.4\arcmin$& 0.25\arcsec\\
6 &  441 & $4.7\arcmin$ & 0.05\arcsec &  517 & $4.6\arcmin$ & 0.05\arcsec & 163 & $4.7\arcmin$ & 0.14\arcsec & 102 & $4.5\arcmin$ & 0.10\arcsec & 147 & $4.5\arcmin$& 0.10\arcsec\\
7 &  795 & $7.7\arcmin$ & 0.05\arcsec & 1119 & $7.6\arcmin$ & 0.05\arcsec & 314 & $7.7\arcmin$ & 0.09\arcsec & 106 & $7.7\arcmin$ & 0.16\arcsec & 191 & $7.7\arcmin$& 0.10\arcsec\\
8 &  376 & $7.7\arcmin$ & 0.08\arcsec &  433 & $7.5\arcmin$ & 0.06\arcsec & 163 & $7.7\arcmin$ & 0.19\arcsec &  64 & $7.5\arcmin$ & 0.17\arcsec &  71 & $7.5\arcmin$& 0.17\arcsec\\
9 &  391 & $4.0\arcmin$ & 0.08\arcsec &  306 & $4.0\arcmin$ & 0.16\arcsec & 122 & $4.0\arcmin$ & 0.28\arcsec &  64 & $3.9\arcmin$ & 0.26\arcsec &  69 & $3.9\arcmin$& 0.30\arcsec\\
J1124--5916$^{*}$\tnote{1}& 4804 & $0.6\arcmin$ &  &  4698 & $0.6\arcmin$ &  & 1961 & $0.5\arcmin$ &  &  1436 & $0.7\arcmin$ &  &  1273 & $0.7\arcmin$ & \\
\hline
\end{tabular}
\begin{tablenotes}
        \footnotesize
        \tablenotetext{\dagger} {The point source's number of counts are from energy band 0.5-7.0 keV.}
        \tablenotetext{*} {The pulsar's number of counts are from energy band 1.2-7.0 keV.}
\end{tablenotes}
\end{center}
\label{tab:counts}
\end{table*}

We investigated this relationship by conducting simulations where we modeled  sources with 5--400 counts and at off-axis angles from 1$\arcmin$--6$\arcmin$.  The results of the simulations are shown in Figure~\ref{fig:positionerror}.  The data follow a roughly power-law relationship between measured position error and surface brightness and off-axis angle, until the surface brightness decreases below $\sim5$ counts arcsec$^{-2}$  and the off-axis angle increases above 4$\arcmin$.  
Sources with the highest surface brightness and smallest off-axis angles have the smallest positional errors. At lower surface brightness values and larger off-axis angles, the power-law relationship begins to break down and the positional errors increase significantly (Figure~\ref{fig:positionerror}). We interpret this to mean that for a point source with lower counts and/or higher off-axis angle, our method will result in a larger error in the measured position. 

We use the results of the simulations shown in Figure~\ref{fig:positionerror} to develop criteria to inform us on the suitability of sources for image registration in the 2006 and 2016 observations. The 2006 observations \ObsID 6677 and \ObsID 6679 are $\sim$ 3 times longer than the 2016 observations (Table~\ref{tab:obslist}). Additionally, the continued accumulation of the ACIS contaminant will reduce the relative number of counts in any one source between 2006 and 2016. Therefore, we apply different criteria for source acceptance between the 2006 and 2016 observations. For the purposes of registration at high precision, we exclude point sources with measured positional errors larger than $0\farcs1$ in the 2006 observations \ObsID 6677 and \ObsID 6679. For the 2006 observation \ObsID 8211, we exclude point sources with measured positional errors larger than $0\farcs15$. In order to include enough sources for proper registration of the 2016 observations, we exclude those sources which have measured positional errors greater than $0\farcs2$.

\subsection{Optical Counterparts} \label{sec:registration}
\begin{figure}[ht!]
\plotone{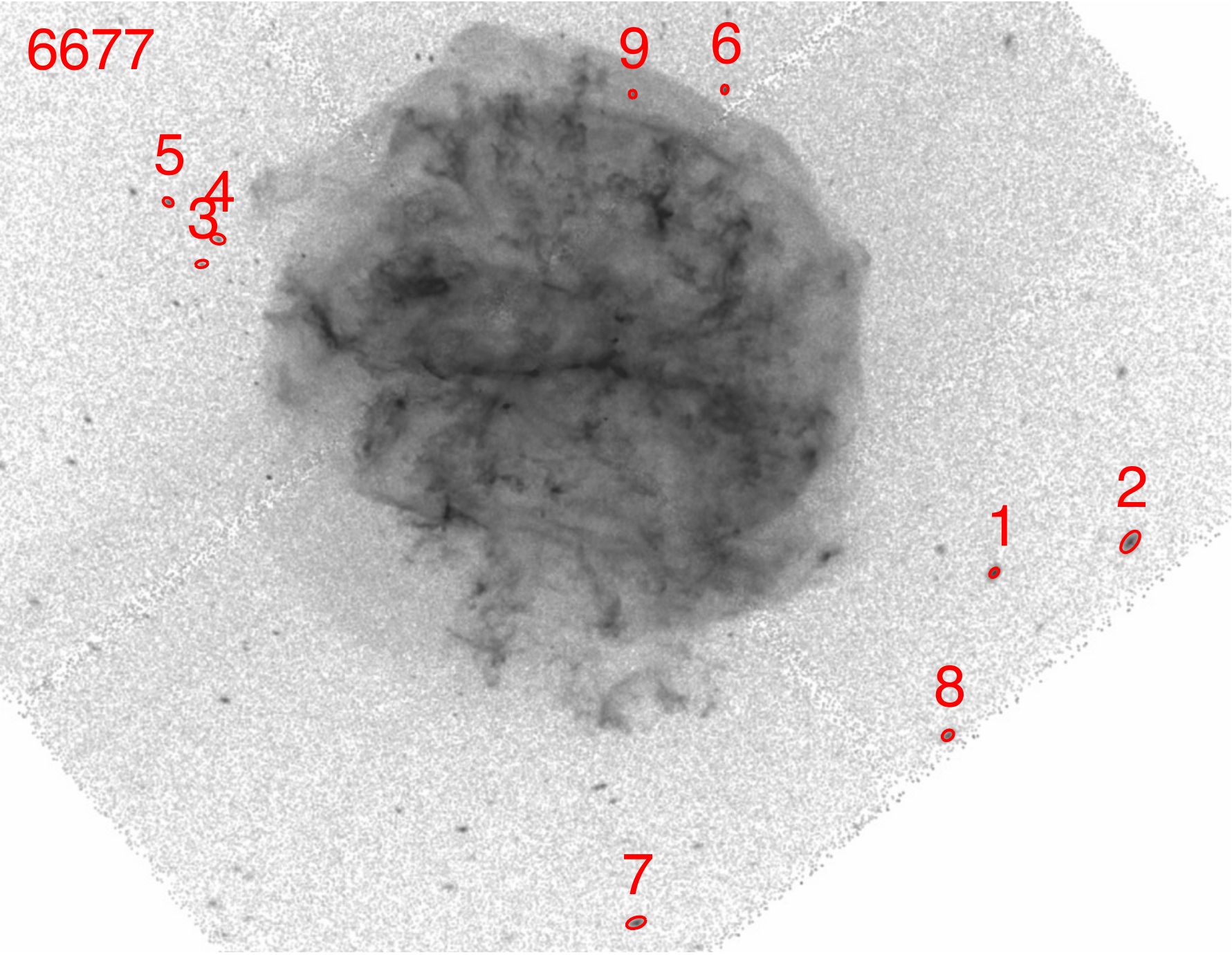}
\caption{The x-ray point sources used to register to optical counterparts from GAIA DR3. See Table~\ref{tab:sourcelist} for details on the individual sources.}
\label{fig:pointsource}
\end{figure}

As noted in \S~\ref{sec:tools}, we used \texttt{wavdetect} to detect X-ray point sources for all four observations. We only use those point sources which are detected in all four observations and which meet our criteria laid out at the end of \S~\ref{sec:pointsource}. For each X-ray point source, we searched GAIA DR3 to find an optical counterpart within $1.5 \arcsec$ radius of the X-ray source. We identified nine optical point sources which match our detected X-ray sources (Figure~\ref{fig:pointsource}). 

\begin{figure*}
    \centering
    \includegraphics[width=0.9\textwidth]{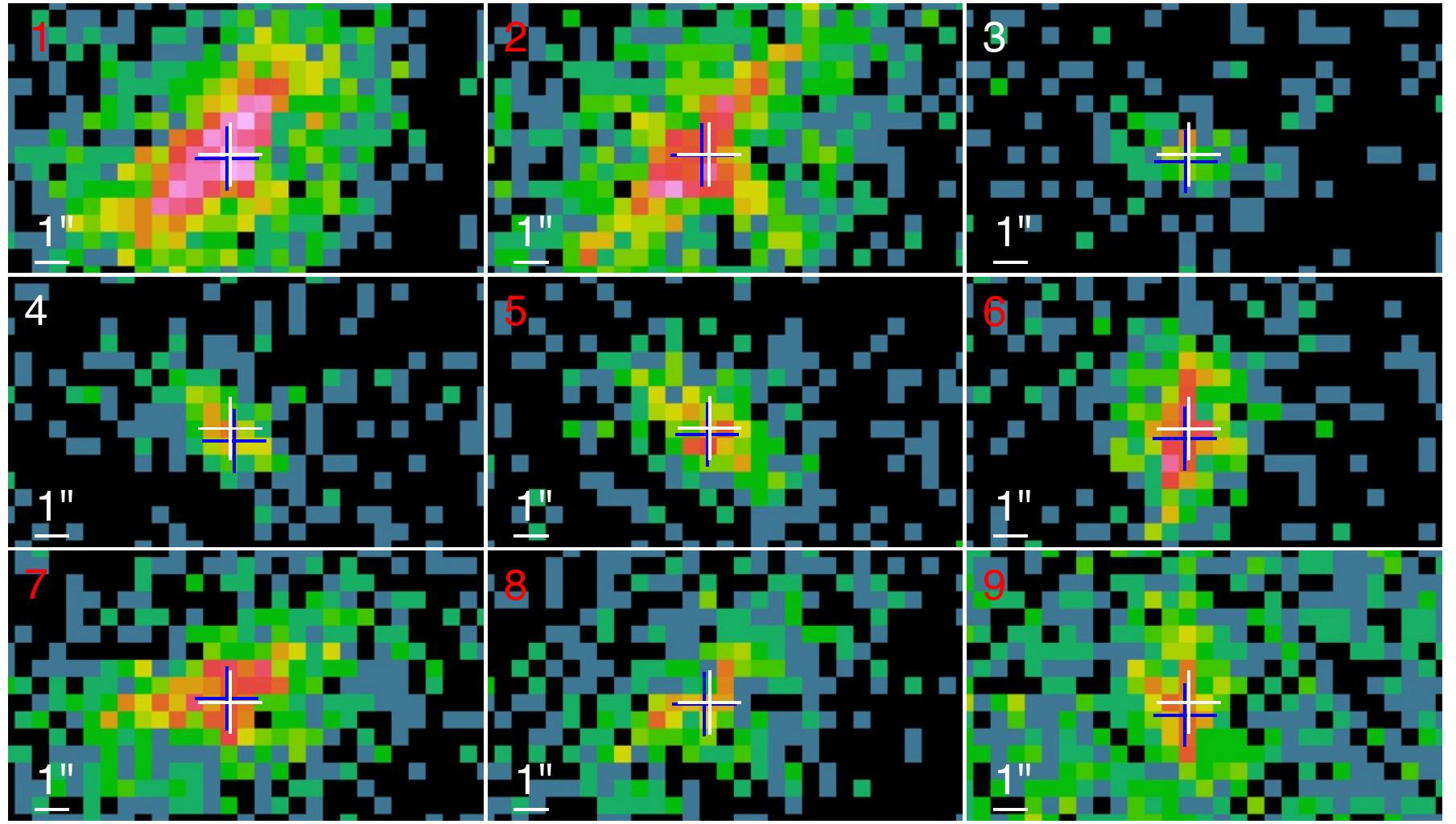}
    \caption{The GAIA point source position (blue crosses) at the epoch of the start time of \ObsID 6677. White crosses indicate the X-ray source positions based on the PSF fit method. The red numbers in the upper left corners identify the sources used in the registration of this obsid, while white numbers indicate that the source was not included in the registration process.\label{fig:source6677}}
\end{figure*}

\begin{figure*}
    \centering
    \includegraphics[width=0.9\textwidth]{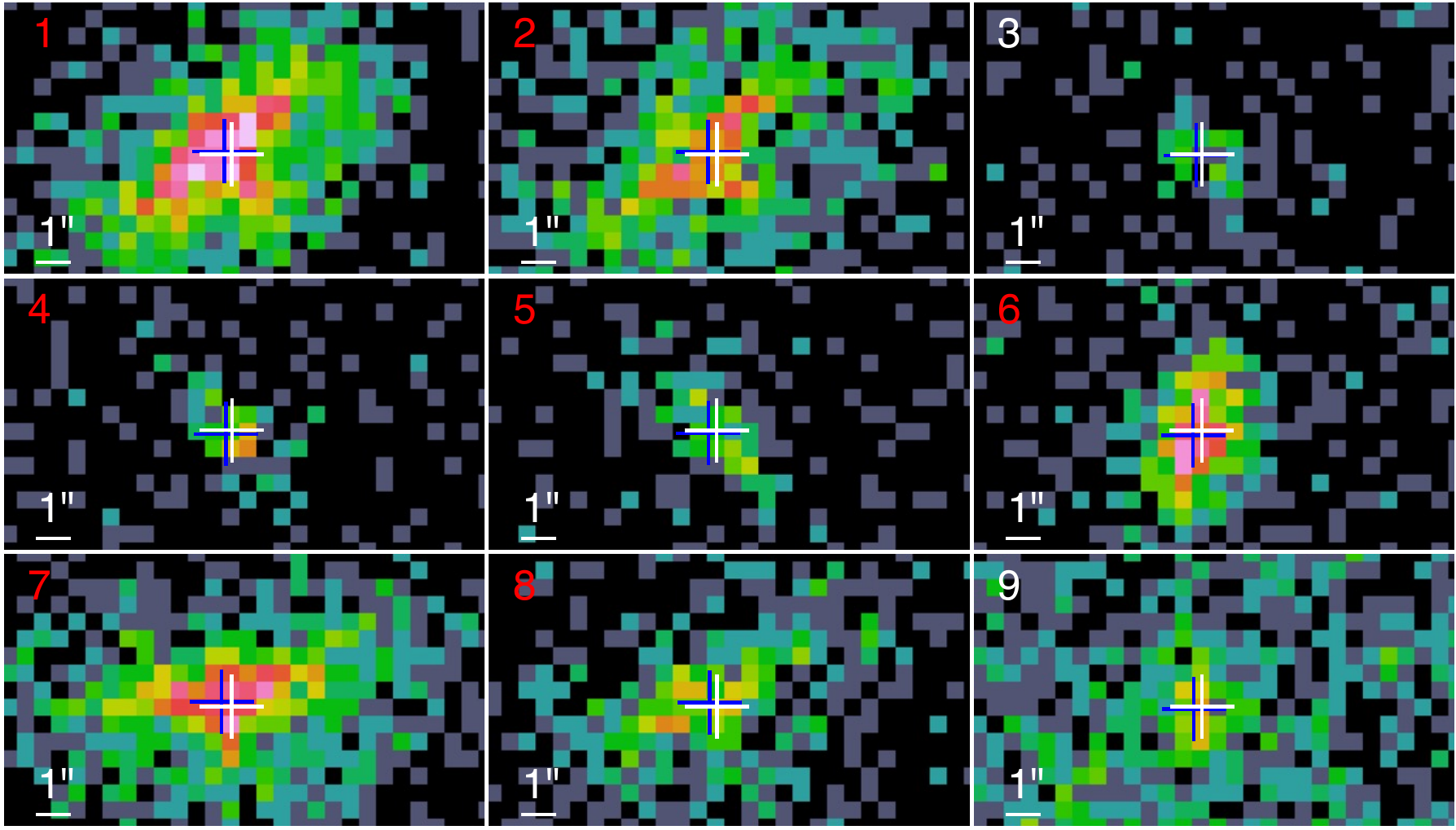}
    \caption{The GAIA point source position (blue cross) at the epoch of the start time of \ObsID 6679.  White crosses indicate the X-ray source positions based on the PSF fit method. The red numbers in the upper left corners identify the sources used in the registration of this obsid, while white numbers indicate that the source was not included in the registration process.\label{fig:source6679}}
\end{figure*}

\begin{figure*}
    \centering
    \includegraphics[width=0.9\textwidth]{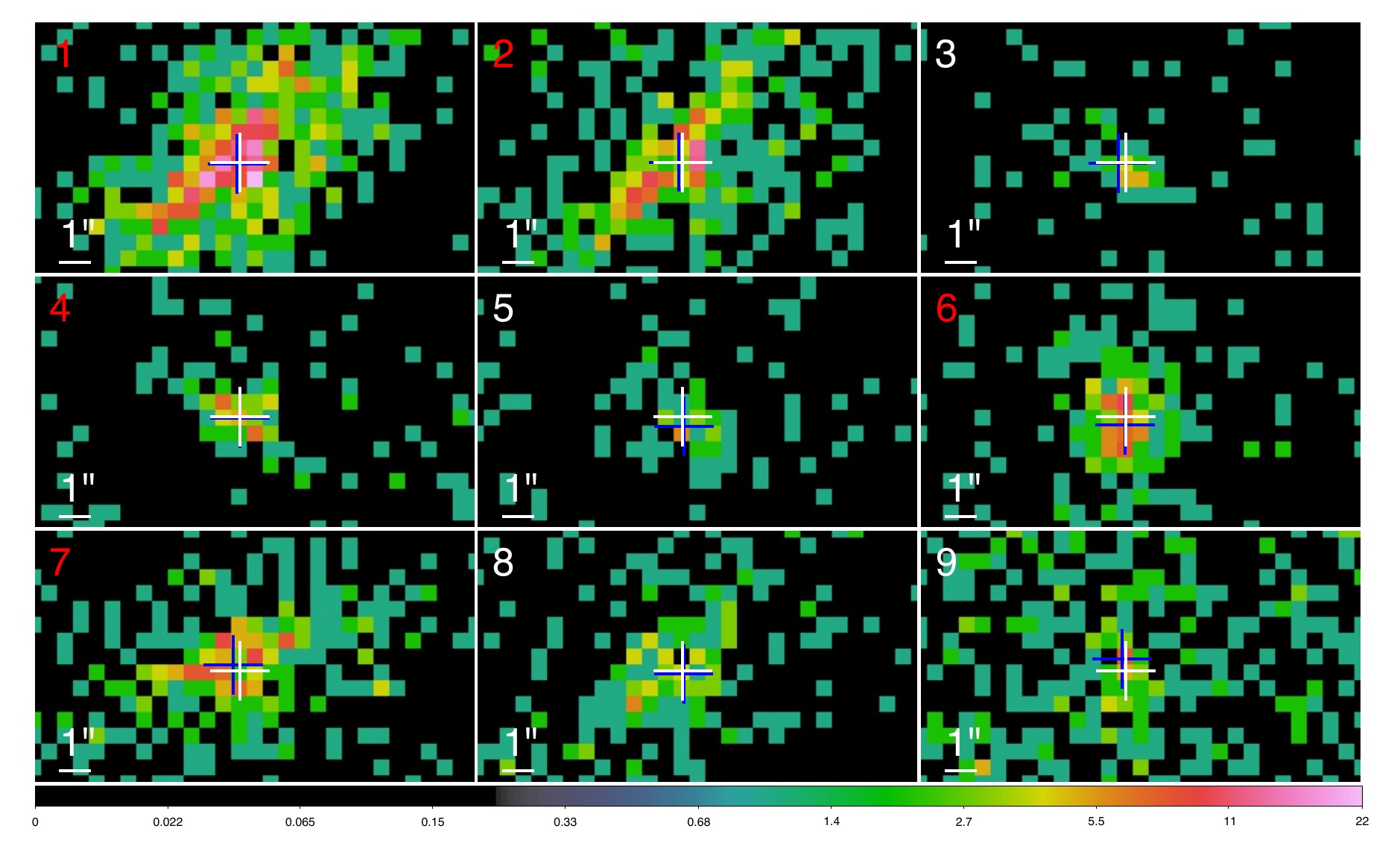}
    \caption{The GAIA point source position (blue cross) at the epoch of the start time of \ObsID 8221.  White crosses indicate the X-ray source positions based on the PSF fit method. The red numbers in the upper left corners identify the sources used in the registration of this obsid, while white numbers indicate that the source was not included in the registration process.\label{fig:source8221}}
\end{figure*}

\begin{figure*}
    \centering
    \includegraphics[width=0.9\textwidth]{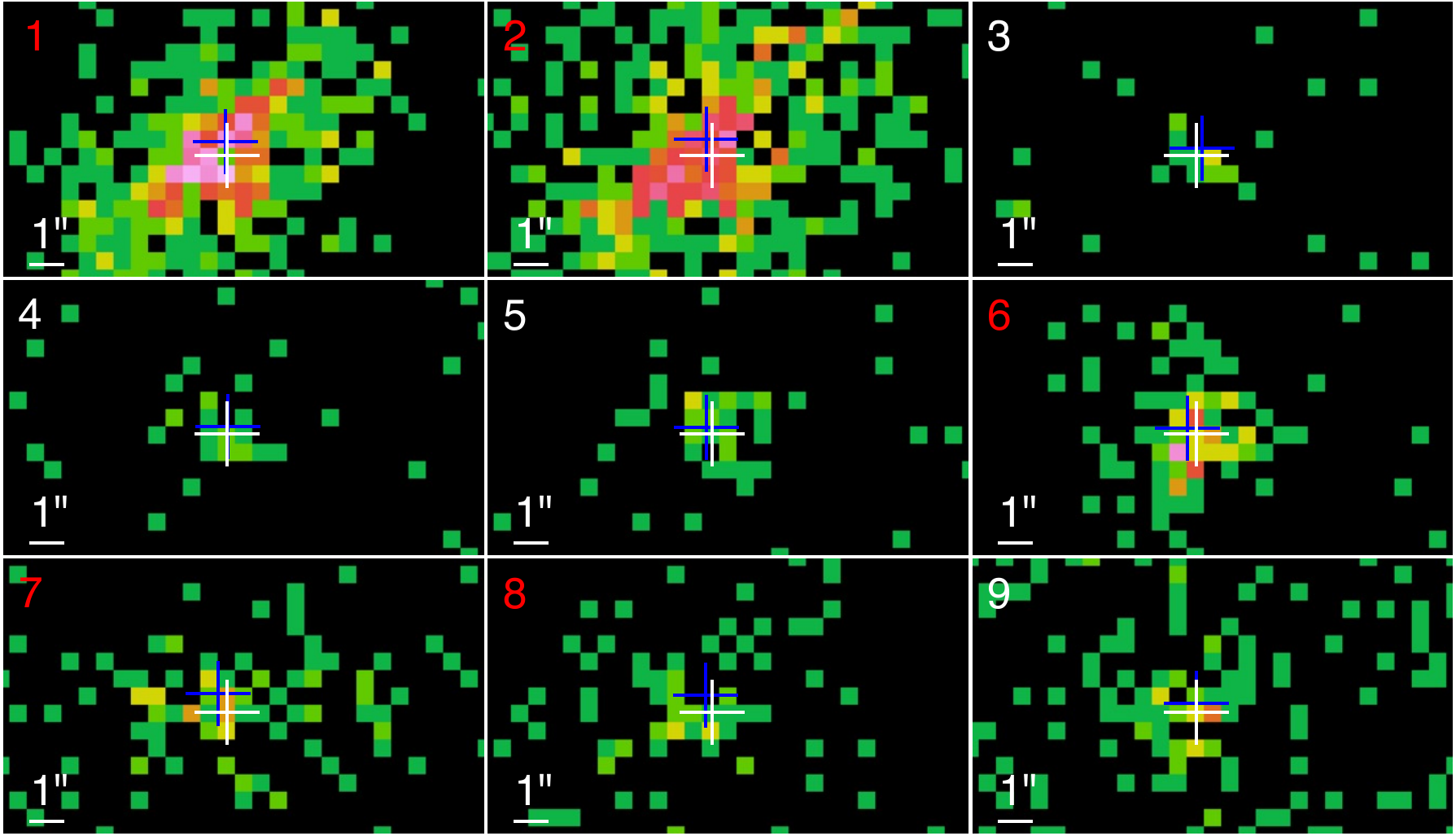}
    \caption{The GAIA point source position (blue cross) at the epoch of the start time of \ObsID 19892.  White crosses indicate the X-ray source positions based on the PSF fit method. The red numbers in the upper left corners identify the sources used in the registration of this obsid, while white numbers indicate that the source was not included in the registration process.\label{fig:source19892}
    }
\end{figure*}

\begin{figure*}
    \centering
    \includegraphics[width=0.9\textwidth]{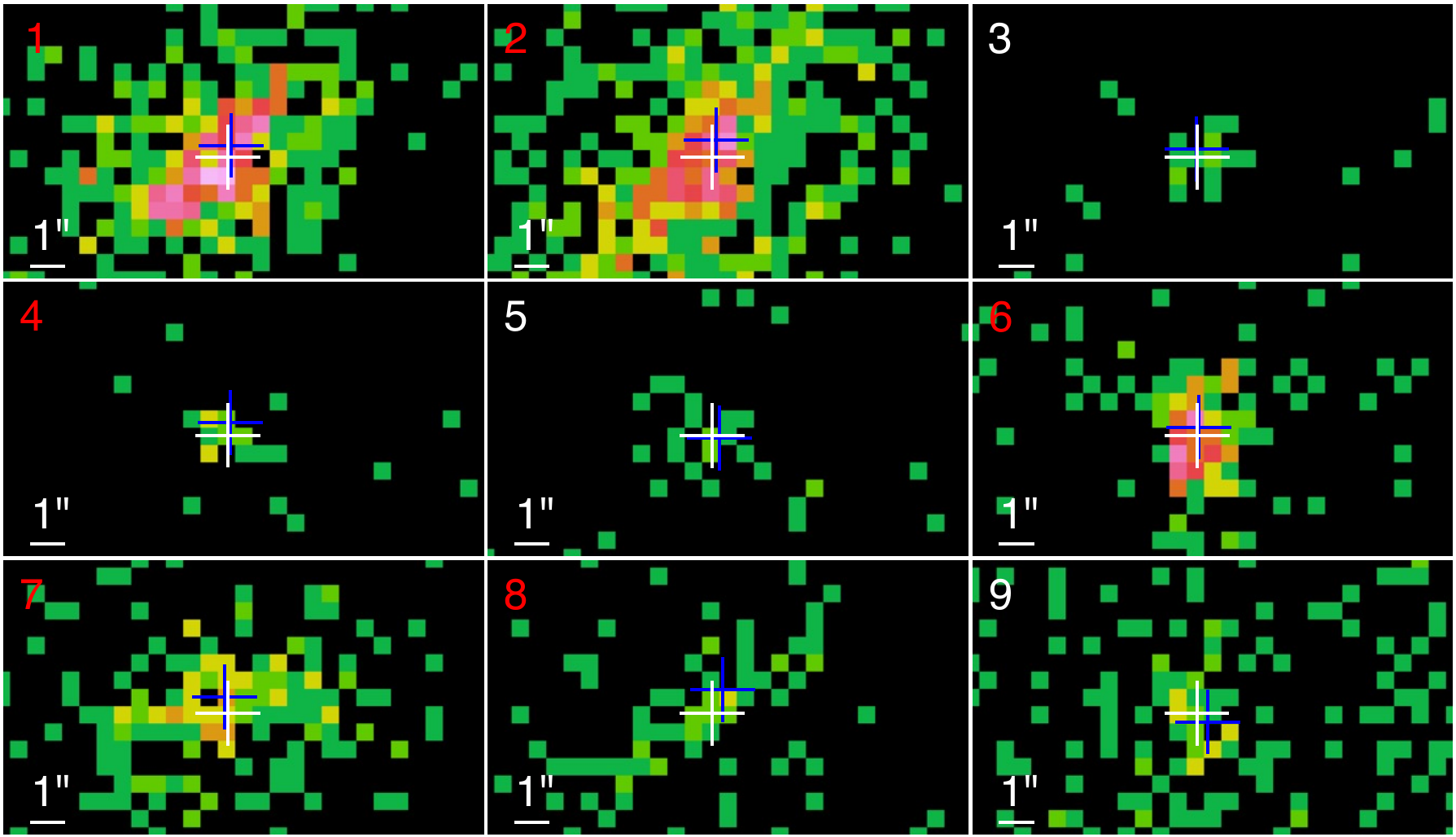}
    \caption{The GAIA point source position (blue cross) at the epoch of the start time of \ObsID 19899.  White crosses indicate the X-ray source positions based on the PSF fit method. The red numbers in the upper left corners identify the sources used in the registration of this obsid, while white numbers indicate that the source was not included in the registration process
    .
    \label{fig:source19899}}
\end{figure*}

Table~\ref{tab:sourcelist} lists the matched GAIA point sources together with coordinates and proper motions. The RA and Dec columns are the values for epoch $2016.0$. We precessed the RA and Dec values of the GAIA sources using the proper motions to the epoch of the start date of each X-ray observation. Accordingly the error of the RA and Dec will be corrected using the error of the proper motion and the time baseline between 2016 and the start dates of the observations as listed in Table~\ref{tab:obslist}. The nine point sources with optical counterparts are shown in Figures~\ref{fig:source6677} -- ~\ref{fig:source19899}, where we plot the GAIA DR3 positions and the X-ray point source fitted positions.

The corrected coordinates of an optical point source are:
\begin{equation}\label{eq:position}
 \begin{pmatrix}
 \alpha\\
 \delta
 \end{pmatrix}
 =
 \begin{pmatrix}
 \alpha_{2016}\\
 \delta_{2016}
 \end{pmatrix}
 +
 \begin{pmatrix}
 v_{\alpha}\cdot \cos(\delta_{2016})\\
 v_{\delta}
 \end{pmatrix} 
\Delta t 
\end{equation}
\noindent Here, $(\alpha, \delta)$ are the corrected RA and Dec values, $(\alpha_{2016}, \delta_{2016})$ are the coordinates of the optical point sources at epoch 2016.0. $v_{\alpha}$, $v_{\delta}$ are the proper motions in Right Ascension and Declination, respectively. The corrected error of RA and Dec value is:
\begin{equation}\label{eq:positionerr}
 \begin{pmatrix}
 \sigma_{\alpha}^{2}\\
 \sigma_{\delta}^{2}
 \end{pmatrix}
 =
 \begin{pmatrix}
 \sigma_{\alpha_{2016}}^{2}\\
 \sigma_{\delta_{2016}}^{2}
 \end{pmatrix}
 +
 \begin{pmatrix}
 \sigma_{v_{\alpha}}^{2} \cdot \cos^{2}(\delta_{2016})\\
 \sigma_{v_{\delta}}^{2}
 \end{pmatrix} 
\Delta t^{2}
\end{equation}

\noindent In Eq~\ref{eq:positionerr}, $(\sigma_{\alpha},\sigma_{\delta})$ are the corrected errors of RA and Dec. $(\sigma_{\alpha_{2016}},\sigma_{\delta_{2016}})$ are the errors of RA and Dec of the optical point sources at epoch 2016.0. $(\sigma_{v_{\alpha}},\sigma_{v_{\delta}})$ are the errors of proper motion in RA and Dec.

\begin{table*}[htb]
\label{tbl:gaia.dr3.point.sources.2016}
\caption{Gaia DR 3 point sources (Epoch 2016)}
\begin{center}
\label{tab:sourcelist}
\begin{tabular}{ l|l c l c c c c c c}
\hline\hline
Source & RA & $\mathrm{RA_{error}}$ & Dec & $\mathrm{Dec_{error}}$ & PMRA  & $\mathrm{PMRA_{error}}$ & PMDec & $\mathrm{PMDec_{error}}$ \\
 ID & $\alpha$ deg & $\sigma_{\alpha}$ mas & $\delta$ deg & $\sigma_{\delta}$ mas & $v_{\alpha}$ $\mathrm{mas\, yr^{-1}}$ & $\sigma_{v_{\alpha}}$ $\mathrm{mas\, yr^{-1}}$ & $v_{\delta}$ $\mathrm{mas\, yr^{-1}}$ & $\sigma_{v_{\alpha}}$ $\mathrm{mas\, yr^{-1}}$ \\
\hline
1 & 170.94734977  & 0.41  & -59.30954920 & 0.39 & -7.00 & 0.55 & 3.30 & 0.43\\
2 & 170.88732949  & 2.40  & -59.30245101 & 2.00 & 0.00 & 0.00 & 0.00 & 0.00\\
3 & 171.29742299  & 0.05 & -59.24006532 & 0.04 & -10.09 & 0.05 & 4.36 & 0.04\\
4 & 171.29006194 & 0.04  & -59.234464888 & 0.03 & -6.60 & 0.05 & 1.90 & 0.04\\
5 & 171.31207660 & 0.03  & -59.226011711 & 0.03 & -6.61 & 0.04 & 1.94 & 0.03\\
6 & 171.06676207 & 0.07  & -59.200712270 & 0.07 & -12.77 & 0.09 & 3.26 & 0.08\\
7 & 171.10550819 & 0.01  & -59.388699202 & 0.01 & -21.10 & 0.01 & 4.37 & 0.01\\
8 & 170.96760530 & 0.02  & -59.346279838 & 0.02 & -10.34 & 0.02 & 4.94 & 0.02\\
9 & 171.10721361 & 0.02  & -59.201670503 & 0.02 & -21.53 & 0.02 & 6.82 & 0.02\\
\hline
\end{tabular}
\end{center}
\end{table*}


\section{Registration method}
\label{sec:registration}

Using point sources from each X-ray observation, we can register the observation by shifting the positions of the point sources, $(x,y)$, to match positions of the corresponding GAIA point sources, $(x_{0}, y_{0})$. The shifted positions are:
\begin{equation}
 \begin{pmatrix}
 x^{\prime}\\
 y^{\prime}
 \end{pmatrix}
 =
 \begin{pmatrix}
 a_{11} & a_{12}\\
 a_{21} & a_{22}
 \end{pmatrix}
 \begin{pmatrix}
 x\\
 y
 \end{pmatrix}
 +
 \begin{pmatrix}
 t_{1}\\
 t_{2}
 \end{pmatrix}
\end{equation}

\noindent
The parameters $a_{11}$, $a_{12}$, $a_{21}$, and $a_{22}$ account for the scale factor and rotation angle, while  $t_{1}$, and $t_{2}$ account for the translation in the x and y directions. 
To match the positions of the point sources to those in the reference observation, $(x_{0},y_{0})$, we minimize a loss function which is weighted by the the position errors: 
\begin{equation}
    D=\sum^{n} \left(\left[1+\frac{(x^{\prime}-x_{0})^{2}}{\sigma_{x}^{2}+\sigma_{x_{0}}^{2}}+\frac{(y^{\prime}-y_{0})^{2}}{\sigma_{y}^{2}+\sigma_{y_{0}}^{2}}\right]^\frac{1}{2} - 1 \right)
\end{equation}
The loss function, $D$, is a sum of softened $L_1$ (absolute value) functions. The softened L1 varies quadratically near the minimum, but asymptotically approaches a linear variation with distance from the minimum. This is provided by the \texttt{soft\_l1} option of \texttt{python} routine \texttt{scipy.optimize.least\_squares}.  This loss function is more robust than least-squares or chi-squared since the weighting of outliers approaches a linear rather than quadratic penalty.
\color{black}
Here $n$ is the number of matched point sources, $\sigma_{x}$ and $\sigma_{y}$ are the position error of the X-ray point sources at positions $(x,y)$; $\sigma_{x_{0}}$ and $\sigma_{y_{0}}$ are the position error of the GAIA point sources at positions $(x_{0},y_{0})$. The fitted parameters $a_{11}$, $a_{12}$, $a_{21}$, $a_{22}$, $t_{1}$ and $t_{2}$ will be used by the \ciao tool \texttt{wcs\_update}, to update the aspect solution file and the WCS of each L2 event list for each observation which is registered to the reference GAIA point sources. Finally, the error of the registration method is estimated by the weighted average position residual of the matched point sources of the two observations:

\begin{equation}\label{eq:regis_err}
    r=\frac{\sum^n (d/\sigma_{d}^2)}
           {\sum^n (1/\sigma_{d}^2)}
\end{equation}

\noindent
In Eq~\ref{eq:regis_err}, $d$ is the position residual of the point sources after registration, $\sigma_{d}$ is the combined position error:
\begin{equation}
  \begin{split}
    d &=\sqrt{(x^{\prime}-x_{0})^{2}+(y^{\prime}-y_{0})^{2}}\\
    \sigma_{d}&=\sqrt{\sigma_{x^{\prime}}^{2}+\sigma_{x_{0}}^{2}+\sigma_{y^{\prime}}^{2}+\sigma_{y_{0}}^{2}}
  \end{split}
\end{equation}

\noindent
The registration error of each X-ray observation to the GAIA point sources is shown in Table~\ref{tab:registration}.

\begin{table}[htb]
\caption{Registration error of X-ray data to Gaia point sources.}
\begin{center}
\label{tab:registration}
\begin{tabular}{ l r r r r r}
\hline\hline
ObsID & 6677 & 6679 & 8221 & 19892 & 19899\\
\hline 
n & 7 & 7 & 5 & 5 & 6\\
$r^{*}$& $0.050\arcsec$ & $0.035\arcsec$ & $0.043\arcsec$ & $0.056\arcsec$  & $0.078\arcsec$\\
\hline
\end{tabular}
\begin{tablenotes}
        \footnotesize
        \tablenotetext{*} {Registration error}
\end{tablenotes}
\end{center}
\end{table}

We note here that our ultimate goal is to measure the motion of J1124--5916. Since we are using the positions of the GAIA stars to correct the astrometry of the Chandra observations, we must consider the effects of solar motion and of differential galactic rotation on the measured positions of the GAIA sources and J1124--5916 \citep[e.g.,][]{Halpern2015}. We have considered this effect and found that it results in at most a $0 \farcs 02$ difference in measured positions for our GAIA sources over the $\sim$ 10 year baseline between observations, and the effects of Galactic rotation will not be measurable in the reference frame of our GAIA field stars.

\section{Proper motion of the pulsar} \label{sec:propermotion}
To measure the proper motion of the pulsar, we employed a similar method to that used to determine the positions of the X-ray point sources. After registering the images with the new WCS, for each observation we extracted a $14.76\arcsec \times 14.76\arcsec$ box region around the pulsar. We used the pulsar image from observation \ObsID 6677 as the ``template'' image, and the pulsar image from a later observation, for instance \ObsID 19892, as the data image. In \sherpa the template image was convolved with a 2-D symmetric Gaussian function plus a constant value to make a model image. The Gaussian function is used to reduce the Poisson error among the adjacent pixels, and the constant value is added to account for the difference of the sky background plus diffuse pulsar wind emission level. Even though the pulsar is embedded in a wind nebula and there is shocked ejecta and circumstellar material along the line of sight, it is nevertheless reasonable to assume that the sky background is homogeneous over the small extraction region. 

\begin{figure*}
    \centering
    \includegraphics[width=0.9\textwidth]{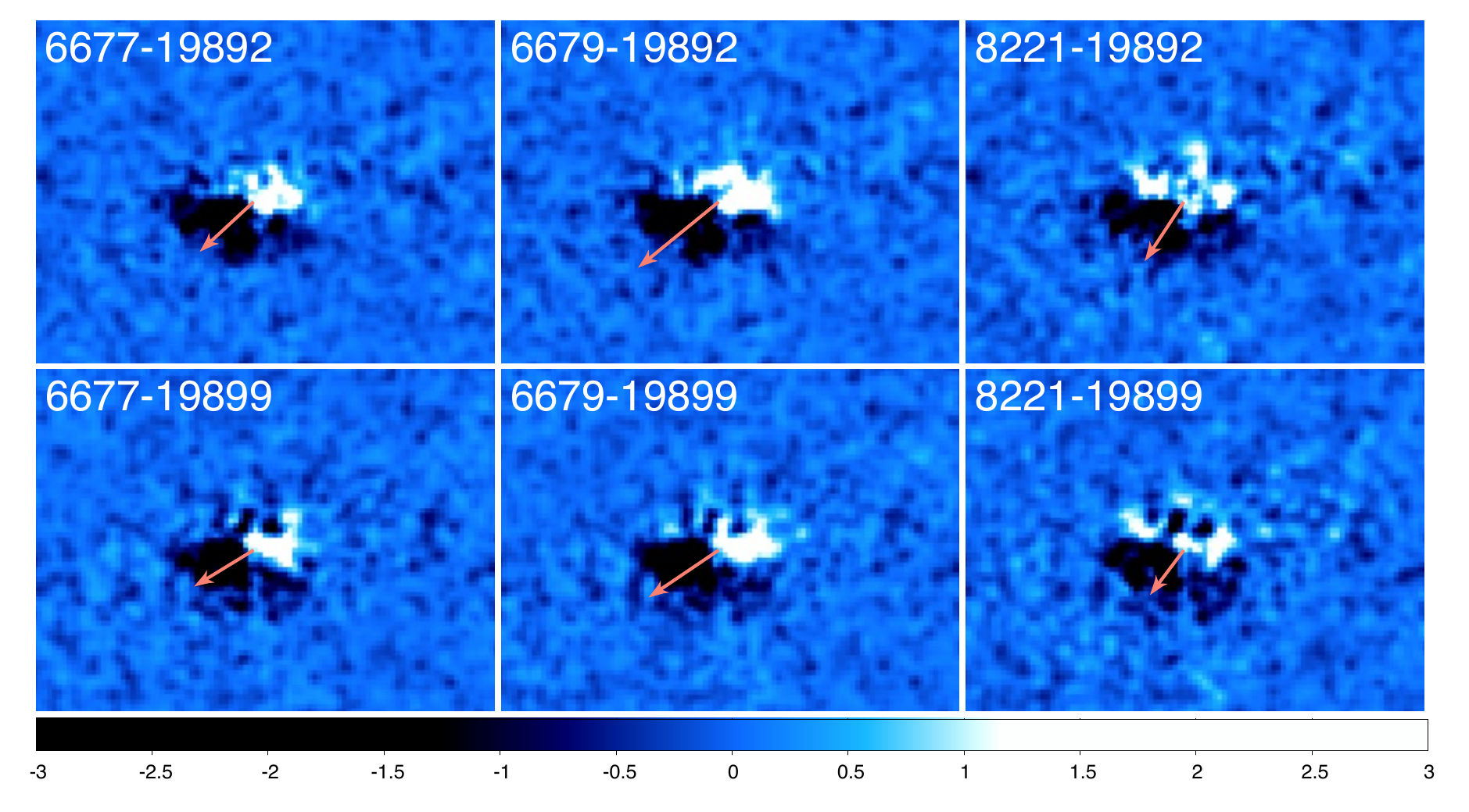}
    \caption{The difference counts images of the pulsar in the energy band 1.2--7.0 keV, in a $9\arcsec \times 7\arcsec$ region, binned to $1/4$ of ACIS sky pixel ($0\farcs123$). The salmon arrows indicate the direction of the proper motion. The salmon-colored arrows show the proper motions scaled by a factor of 10.
    \label{fig:pulsarpm}}
\end{figure*}

In the model image, the center, amplitude and sigma of the Gaussian function and the constant value are free to vary. The C-statistic is again used as the fit statistic. The difference between the fitted center of the Gaussian function $(x,y)$ and that of the center of the pulsar image for which the proper motion is being computed is the absolute shift in the position between the epochs. As in \S~2.2, we use the sherpa \texttt{conf} routine to determine this difference in two dimensions. However, in this case, we use the 2006 observation as our template image, instead of a raytrace of the PSF. The $1 \sigma$ errors of the center $(\sigma_{x},\sigma_{y})$ are calculated by varying the value of $x$ or $y$ along a grid of values while the values of all the other thawed parameters are allowed to float to new best-fit values. The measured shifts and inferred transverse velocities are listed in Table~\ref{tab:propermotion}. In Figure~\ref{fig:pulsarpm} we plot the difference images between the 2006 and 2016 observations; the resulting measurements are presented in Table~\ref{tab:propermotion}. Also shown in Figure~\ref{fig:pulsarpm} are vectors indicating the measured direction of motion of J1124--5916, with lengths that are proportional to the measured shift in the pulsar centroid.

As indicated in Table~\ref{tab:propermotion}, the mean measured positional shift of J1124--5916 is $0.208 \arcsec \pm 0.010 \arcsec$. At a distance of 6.2 kpc to G292.0+1.8, this corresponds to a mean transverse velocity of $\mathrm{612\,km \, s^{-1} \pm 30\,km \, s^{-1}}$. Considering the registration errors of the observations to the GAIA point sources, there is a systematic error which should be added to the proper motion result. To combine the statistical uncertainty and the systematic error, we adopt the multiple imputation method to calculate the total error \citep{Lee2011}. Because the spread in the time deltas between our observations spans the narrow range of 9.96 yr to 10.05 yr, it is reasonable to assume the proper motion does not change within the precision of our measurement over that time period. We have six measurements of the proper motion. The average value of the six measurements is reported as the estimated proper motion. The statistical uncertainty is the average of each individual measurement. The systematic error is from the registration error, and is estimated from the variance of the six measurements. The total error is the combined statistical uncertainty and the systematic error, according to the Equations 4--6 in \citep{Lee2011}. Therefore, the proper motion velocity with the total error is $\mathrm{612\,km \, s^{-1} \pm 152\,km \,s^{-1}}$. 

\begin{table*}[htb]
\caption{Proper motion measured between 2006 and 2016 observations and relative shifts of pulsar position between 2006 and 2016 epochs}
\begin{center}
\label{tab:propermotion}
\begin{tabular}{ l c c c c c c c c}
\hline\hline
ObsID & 6677-19892 & 6677-19899 & 6679-19892 & 6679-19899 & 8221-19892 & 8221-19899 &average & total error\\
\hline 
$\mathrm{\Delta x\,(mas)}$ & $-162 \pm 20 $ & $-176 \pm 21 $ & $-229 \pm 20$ & $-203 \pm 19$& $-125 \pm 21$& $-112 \pm 20$& -168 & 52\\
$\mathrm{\Delta y\,(mas)}$ & $-118\pm 14$ & $-84\pm 14$ & $-158\pm 14$ & $-112\pm 14$ & $-142 \pm 15$ & $-106 \pm 15$ & -120 & 32\\
$\mathrm{\Delta R\,(mas)}$ & $201 \pm 24 $ & $195 \pm 25 $ & $278 \pm 25$ & $232 \pm 24$ & $189 \pm 26$ & $154 \pm 25$ & 208 & 52\\
$\mathrm{\Delta t\,(yrs)}$ & 9.97 & 10.02 & 10.02 & 10.05 & 9.96 & 10.00 & -- & --\\
$\mathrm{v\,(km\,s^{-1})}$ & $592\pm 71 $ & $572 \pm 73 $ & $ 816 \pm 73$ & $679 \pm 70$ & $558 \pm 76$ & $452 \pm 74$ & 612 & 152\\
Position angle & $126 \pm 7$ & $116 \pm 6$ & $125 \pm 5$ & $119 \pm 5$ & $139 \pm 8$ & $133 \pm 9$ & 126 & 17\\
\hline
\end{tabular}
\end{center}
\end{table*}

\section{Discussion} \label{sec:discussion}

\begin{figure*}
\centering
\includegraphics[width=\textwidth]{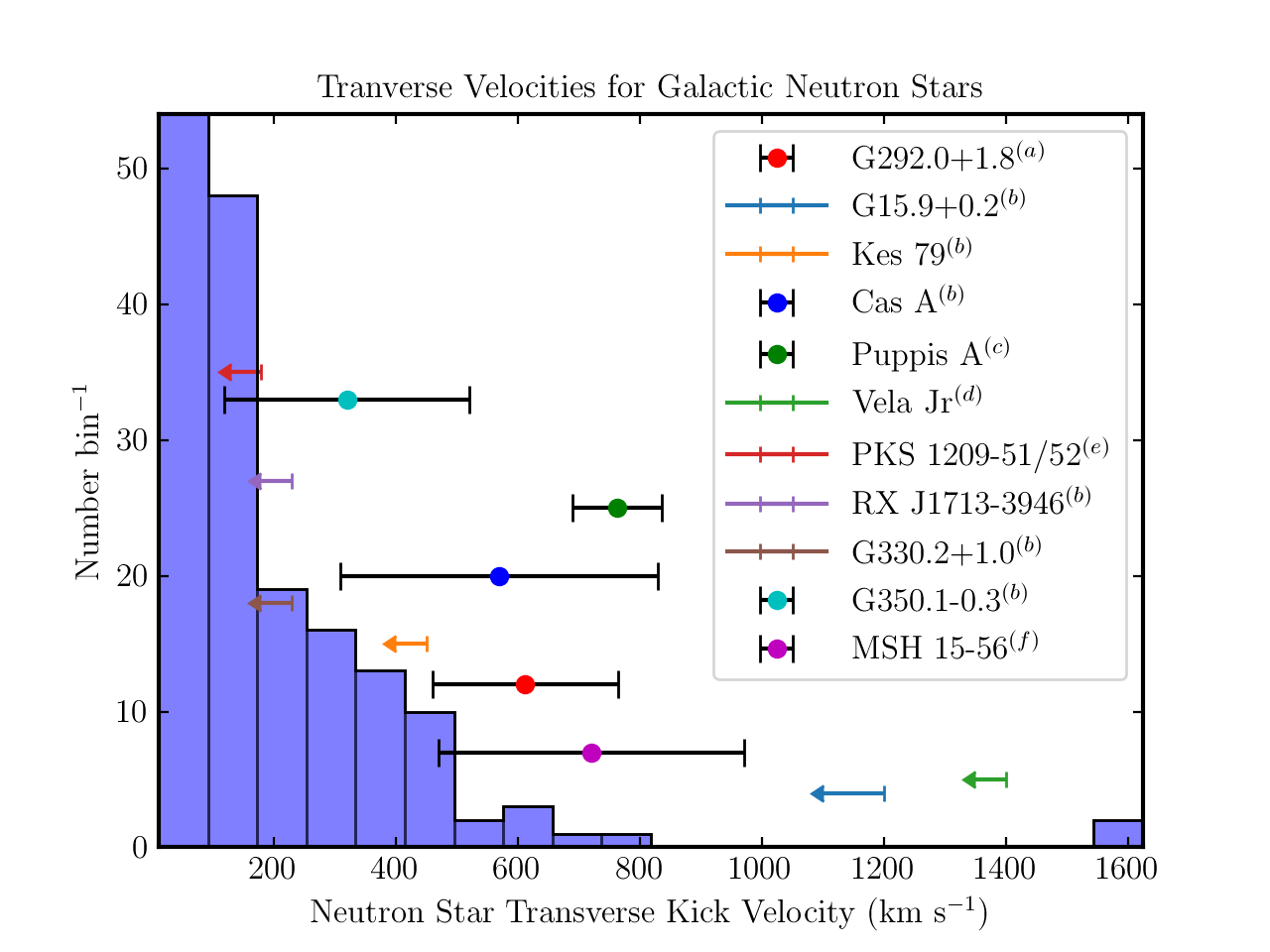}
\caption{Distribution of transverse neutron star kick velocities from \citet{Hobbs2005}. We overplot our measurement as well as  measurements from recent studies. \textsc{References:} (a): This work; (b): \citet{Mayer2021}; (c): \citet{Mayer2020}; (d): \citet{Mignani2007,Mignani2019}; (e): \citet{Halpern2015}; (f): \citet{Temim2017}.} 
\label{fig:nskicks}
\end{figure*}

As mentioned in Section~\ref{sec:propermotion}, we measure a proper motion of 612 km s$^{-1}$ assuming a distance of 6.2 kpc. In Figure~\ref{fig:nskicks}  we plot our measured kick velocity for the neutron star in G292 against the distribution of Galactic neutron stars presented in \citet{Hobbs2005}. We include recent results presented by \citet{Mayer2021} for a sample of central compact objects (CCOs) in SNRs, and the recent measurement of the pulsar proper motion in MSH 15--56 \citep{Temim2017}. 

As seen in Figure~\ref{fig:nskicks}, the proper motion of the neutron star in G292 lies at the higher end of the distribution of kick velocities for Galactic neutron stars, and has a transverse kick velocity similar to the pulsar located in MSH 15-56 \citep{Temim2017}. \citet{Wongwathanarat2013} noted that some neutrino-driven supernova models can impart kick velocities in excess of 600 km s$^{-1}$ within the first few seconds of core-collapse, but that study only considered 20 models, with a limited range of progenitor masses. Interestingly, Figure~8 of \citet{Wongwathanarat2013} suggests that the 20 M$_{\sun}$ progenitor model did not impart a large kick velocity on the neutron star. The progenitor mass for G292 is poorly constrained, with an estimate between 13--30 M$_{\sun}$ \citep{Bhalerao2019}, though a recent study by \citet{Jacovich2021} suggests a progenitor mass of $\sim$ 22~M$_{\sun}$. In this sense and in light of this new result, additional high fidelity simulations may be required in order to connect the measured kick velocity back to supernova explosion models.

We also consider whether the kick velocity arises from anisotropic neutrino emission. \citet{Burrows2007} parametrized the degree of anisotropy  as $\sin{(i)}$ and derived an expression for the kick velocity as: 
\begin{equation}
    V_k \sim 1000 \left(\frac{E_{SN}}{10^{51} \, \mathrm{erg}}\right) \sin{(i)} \, \mathrm{km \ s^{-1}}\, ,
\end{equation}

\noindent
For our measured value of 612 km s$^{-1}$ we require an explosion energy of $\sim$ 10$^{51}$ erg and a high degree of anisotropy is required ($\sin{(i)}$ $\sim$ 1). In studying the X-ray properties of the SNR shock, \citet{Lee2010} assumed an explosion energy of 0.5--1.0$\times$10$^{51}$ erg. The models of \citet{Jacovich2021} were tuned to match the $^{56}$Ni yields of 1D explosion models, with explosion energies of $\sim$ 8$\times$10$^{50}$ erg, which is bracketed by the assumed energies in \citet{Lee2010}. In order to match the blastwave kinematics, a lower explosion energy requires a lower ejecta mass. The cradle-to-grave models of \citet{Jacovich2021} give an ejecta mass of $\sim$ 10M$_{\sun}$, while those of \citet{Lee2010} range from $\sim$ 5 -- 20 M$_{\sun}$. 20 M$_{\sun}$ of ejecta seems unlikely, ruling out an explosion energy of 10$^{51}$ erg, unless the density of the circumstellar environment was considerably higher than what was assumed in \citet{Lee2010}. In light of this, we choose explosion energies of 0.5--0.8$\times$10$^{51}$ erg. This gives an asymmetry parameter $\sin{(i)}$ $\gtrsim$ 0.8. This degree of anisotropy is considerably higher than what is found for the neutron star in Puppis A \citep[e.g., $ \sin{(i)} $ $\sim$ 0.7;][]{Becker2012}, suggesting that the lower explosion energies required to explain the X-ray properties are only feasible if the explosion that imparted the high kick velocity on the G292 neutron star was highly asymmetric. 

\begin{figure*}
    \centering
    \includegraphics[width=\textwidth]{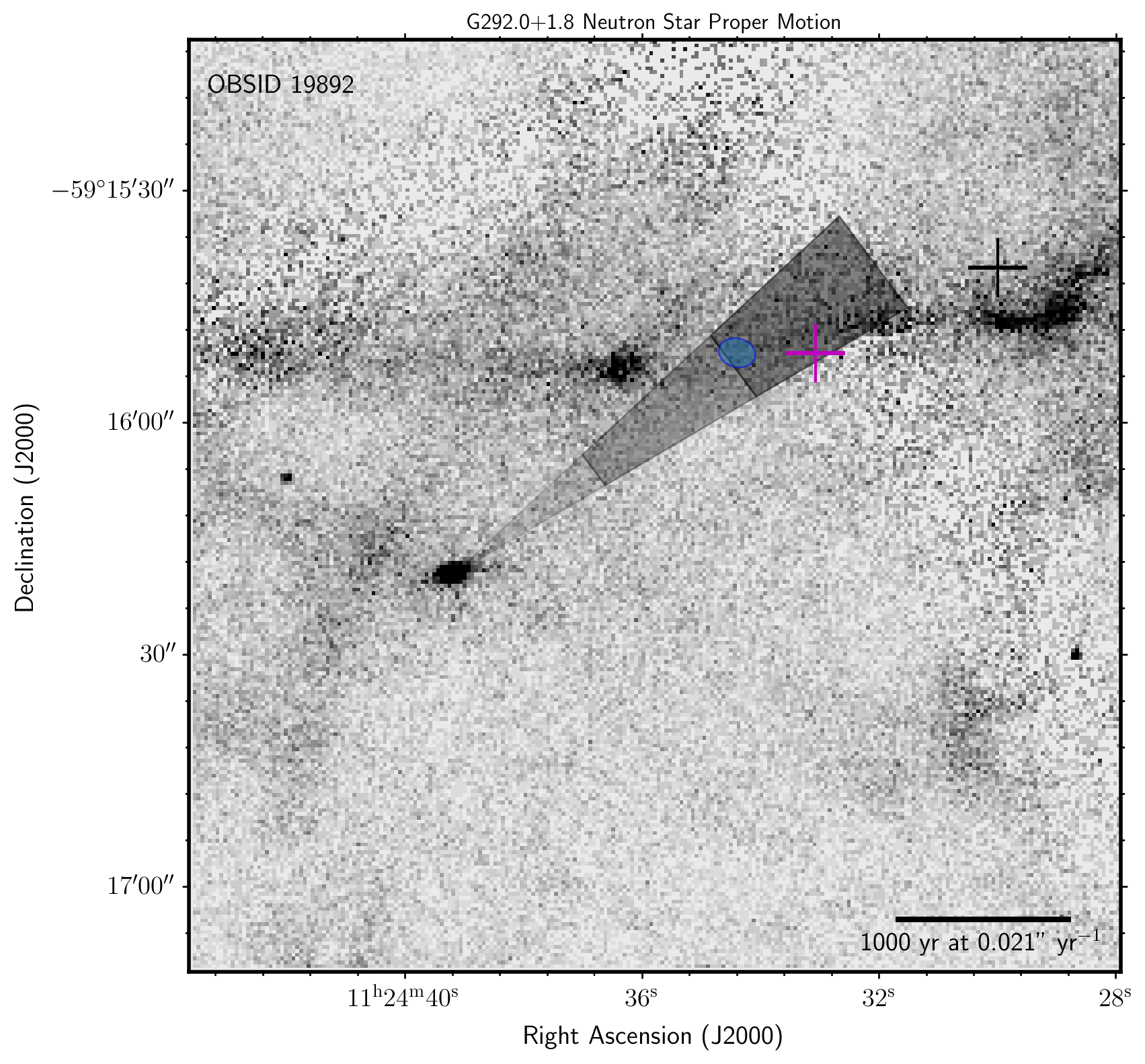}
    \caption{Projection of G292 neutron star back to the center of expansion. The opening angle corresponds to the range of position angle (including the error on that angle). We assume the mean proper motion of $0\farcs021$ yr$^{-1}$. The blue shaded ellipse corresponds to the center of expansion from proper motion measurements of optical ejecta \citep{Winkler2009}. The magenta cross corresponds to the geometrical center of the X-ray emission, while the black cross corresponds to the center of mass for the distribution of intermediate mass elements \citep{Katsuda2018}. The length of each shaded segment corresponds to 1000 years of motion of the neutron star, assuming no deceleration.}
    \label{fig:ns-pm}
\end{figure*}

We are now in a position to compare the measured proper motion to the center of expansion as determined from optical observations \citep{Winkler2009} and from the distribution of intermediate mass elements in the shocked ejecta \citep{HollandAshford2017,Katsuda2018}. In Figure~\ref{fig:ns-pm} we project the motion of the neutron star back 3000 years, using our mean proper motion of $0\farcs021$ yr$^{-1}$. We also mark the geometric center of expansion and the center of mass for intermediate mass elements as measured by \citet{Katsuda2018}. Using the range of position errors, we find good agreement with our direction of motion and the center of expansion measured from proper motions of optical knots \citep{Winkler2009}. 

G292 is quoted as being 2990$\pm$60 years old \citep{Winkler2009}. As seen in Figure~\ref{fig:ns-pm}, our mean proper motion implies an age much closer to 2000 years old, suggesting that G292 might be the remnant of an historical supernova. Unfortunately, the low declination of G292 means that a historical companion to G292 would be below the horizon to Europe, China, and the Middle East \citep{Clark1976}. The closest historical SN is that of AD 185, but a reconstruction of the historical record places all candidate SNRs for that event in the range of 310--320$\degr$ in Galactic longitude \citep{Stephenson2002}. 

\begin{figure*}
    \centering
    \includegraphics[width=0.5\textwidth]{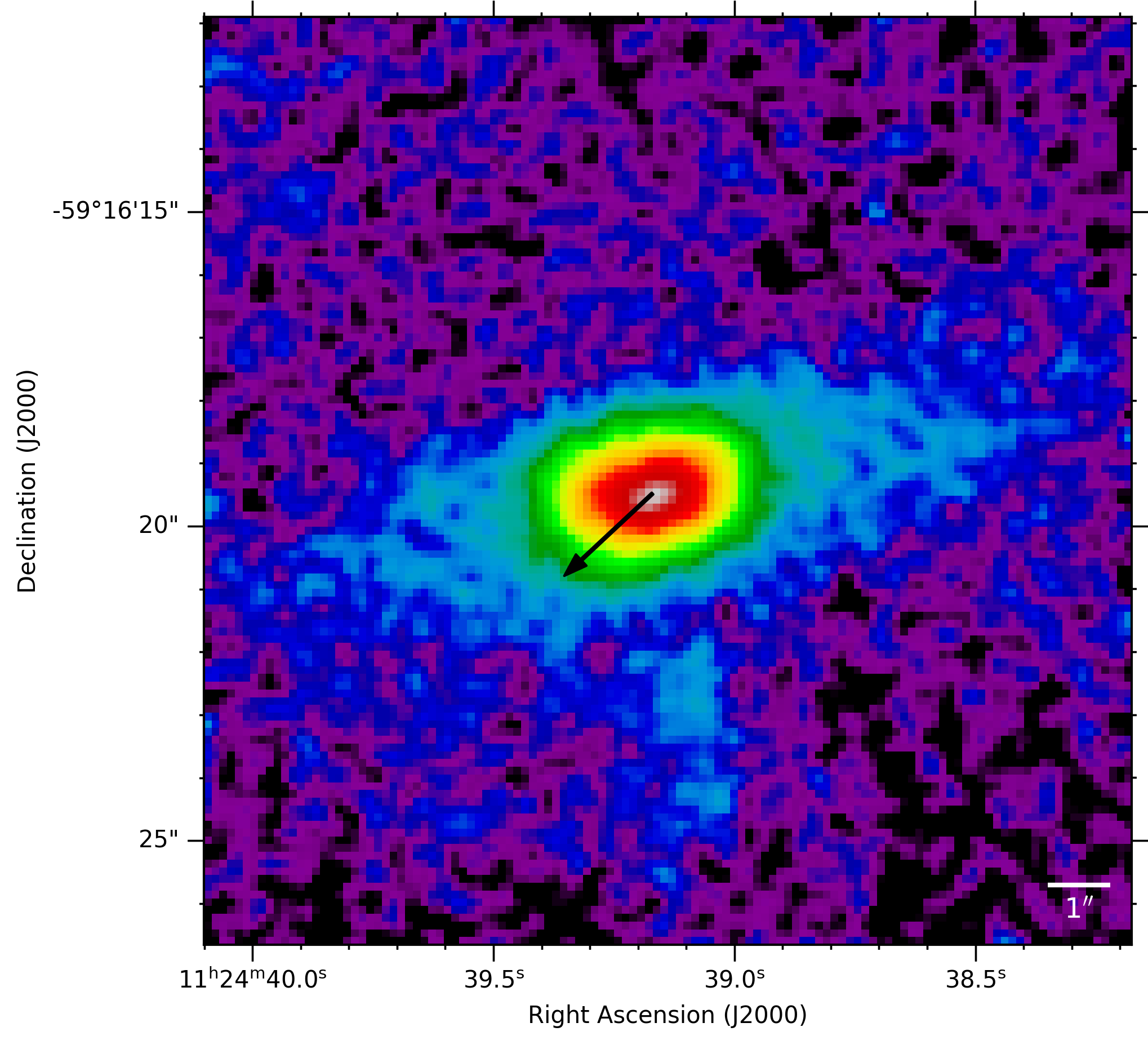}
    \caption{The pulsar and pulsar wind nebula image stacked from all 5 observations registered to the observation \ObsID 6677. The bin size is 1/4 sky pixel ($0.123\arcsec$). The black arrow shows the proper motion direction of the pulsar; the length of the arrow is 10 times the measured shift of the pulsar. The image has been smoothed with a Gaussian kernel with a $\sigma$ of 1 bin and diameter of 5 bins.}
    \label{fig:ns-rot}
\end{figure*}

Lastly, we briefly comment on the kick direction with respect to the apparent rotation of the pulsar. In Figure~\ref{fig:ns-rot}, we show a zoomed in image of the pulsar wind nebula, with the direction of motion indicated. \citet{Park2007} first noted in a deep \textit{Chandra} observation of G292.0+1.8 the torus and jet structure in the PWN in high detail, with the jet appearing to be aligned along the north--south axis \citep[see inset of Figure~1;][]{Park2007}. The nearly north-south alignment of the jet is misaligned from the kick direction by $\sim$ 45$\degr$. This observed misalignment is consistent with recent results which demonstrated a seemingly random distribution of spin--kick alignments in 3D simulations of core-collapse supernovae \citep{Janka2022}, though at odds with observations of spin--kick alignment in other systems \citep[e.g.,][]{ng2007,noutsos2012,yao2021}.

\section{Conclusions} \label{sec:conclusions}
We measure a position shift of the pulsar in supernova remnant G292.2+1.8 of $0\farcs21 \pm 0\farcs05$ over a 10 year observation baseline. At a distance of $\sim$ 6.2 kpc, this corresponds to a transverse velocity of: $\mathrm{612\,km \, s^{-1} \pm 152\,km \, s^{-1}}$. The measured velocity of the pulsar is $\sim$ 30\% higher than the $\sim$ 450 km s$^{-1}$ which was determined by comparing the position of the pulsar to that of the optical center of expansion inferred from kinematic fits to the motion of O-rich ejecta knots which also give an age of $\lesssim$ 3000 yr \citep{Winkler2009}.

We compute the degree of neutrino anisotropy which is required to impart a kick velocity of 600 km s$^{-1}$, and find that an extreme degree of anisotropy is required to explain the high velocity, unless the explosion energy was considerably higher than what is typically assumed. A high explosion energy is inconsistent with other observables in G292. We therefore conclude that the neutron star kick in G292 has a hydrodynamic origin.

Lastly our new measured velocity suggests an age $\sim$ two-thirds of what is typically assumed. For a 2000 year old SNR, we explore the possibility that G292 is the result of a historical supernova. However, a literature search notes that the Galactic longitude of G292 is below the horizon for most northern hemisphere civilizations that might have observed it. The closest candidate is SN AD 185, which is still located several degrees in Galactic latitude and longitude from G292. There is no record in the literature of a SN being observed in the southern hemisphere in the direction of G292.

\begin{acknowledgments}
T.J.G, D.J.P, and P.P.P acknowledge support under NASA Contract NAS8-03060. X.L. acknowledges support from CXC grants SP8-19002X and GO9-20068X, and NASA grant 80NSSC18K0988.
\end{acknowledgments}

%

\vspace{5mm}
\facilities{\chandra, GAIA}


\software{\ciao\citep[v4.13;][]{fruscione2006}, \marx\citep[v5.5.1;][]{davis2012}, \saotrace\citep[v2.05;][]{jerius2004}, \sherpa\citep{freeman2001,doe2007,burke2020}, \textsc{AstroPy}\citep{astropy:2013,astropy:2018}}



\appendix

\section{A Determination of the Sytematic Error on Registration with GAIA Sources}

In order estimate the systematic error on the registration of our \chandra data with the {\it GAIA} sources and to provide a sanity check on our method, we compared one observation from one epoch (either 2006 or 2016) to another observation from the same epoch and searched for a shift in position.  We generated a difference image of the pulsar using the registered observations \ObsID 6677 and 6679, which are observed within one week of each other, and therefore the pulsar should have no shift in position.

\begin{figure*}
    \centering
    \includegraphics[width=0.9\textwidth]{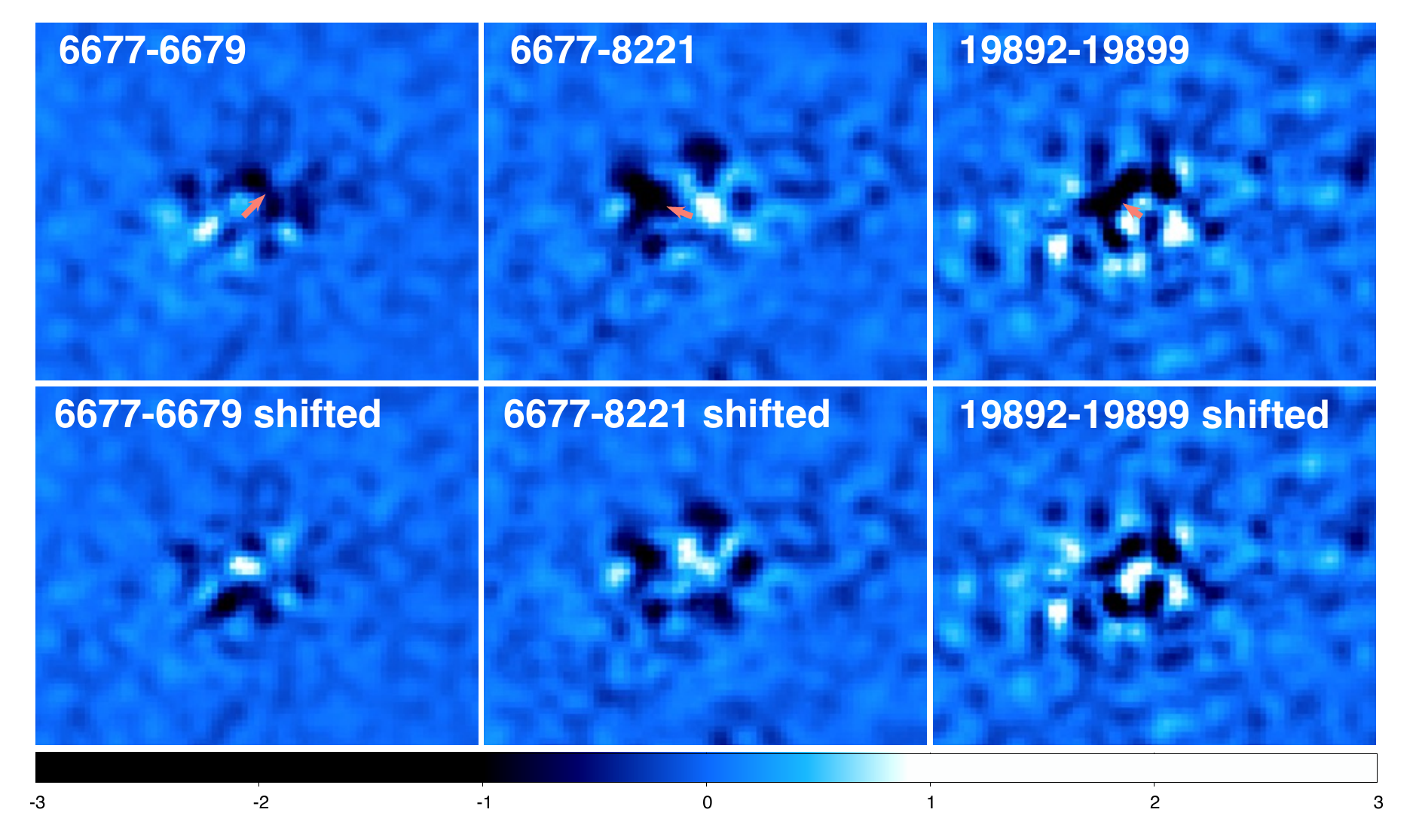}
    \caption{The difference counts image of the pulsar of observations \ObsID 6677 and \ObsID 6679, \ObsID 6677 and \ObsID 8221, \ObsID 19892 and \ObsID 19899 in energy band 1.2--7.0 keV, the bin size is $1/4$ of ACIS sky pixel ($0.123\arcsec$). The images have been smoothed in DS9 with a Gaussian kernel. The $\sigma$ and radius are set at 1.5 bins and 3 bins, respectively. Upper: the difference image after registration. Lower: the difference image corrected the shifted measured using our method for measuring proper motion. \label{fig:diffshift}}
\end{figure*}

The upper left panel of Figure~\ref{fig:diffshift} shows the difference image of the pulsar by subtracting the image of \ObsID 6679 from the image of \ObsID 6677. It seems there is a slight shift in the southeast to northwest direction. Applying our proper motion measurement method, we measured a $0\farcs064 \pm 0\farcs025$ shift listed in Table~\ref{tab:sanity}. This value is consistent with the registration error of \ObsID 6677 and 6679 reported in \S~2. The red arrow shows the direction of the measured shift. In order to highlight the shift, the length of the arrow has been artificially increased by a factor of 10. The bottom left panel of Figure~\ref{fig:diffshift} shows the difference image after correcting the measured shift of $0\farcs064$. The random distribution of white and black pixels in the difference image indicates that there is no obvious shift compared to the distribution in Figure~\ref{fig:pulsarpm} which shows a clear shift in the pulsar position.  This analysis is repeated for 
\ObsID 6677 and 8221 and \ObsID 19892 and 19899 
and the results are shown in the center and right panels of  Figure~\ref{fig:diffshift} and in Table~\ref{tab:sanity}.

The measured shift between \ObsID 6677 and 8221 is $0\farcs056 \pm 0\farcs025$ and the measured shift between \ObsID 19892 and 19899 is $0\farcs047 \pm 0\farcs026$, also consistent with our estimate of the registration error. The statistical fluctuations in the 2016 data are larger given the shorter exposure times of those observations.

According to this analysis the systematic error in the registration of the \chandra data with the {\it GAIA} sources is $0\farcs047-0\farcs064$ which is comparable to our estimate of the uncertainty in the {\it GAIA} registration of $0\farcs035-0\farcs078$ listed in Table~\ref{tab:registration}. Both of these uncertainties are significantly smaller than the measured positional change of the pulsar of $0\farcs21$.

\begin{table*}[htb]
\caption{Relative shifts of pulsar position between 2006 and 2016 epochs}
\begin{center}
\label{tab:sanity}
\begin{tabular}{ l c c c c c}
\hline\hline
ObsID  & 6677-6679 & 6677-8221 & 19892-19899 & average & total error\\
\hline 
$\mathrm{\Delta x\,(mas)}$ & $45 \pm 21 $ & $-52 \pm 21 $ & $-27 \pm 21 $ & $-11$ & 62\\
$\mathrm{\Delta y\,(mas)}$ & $45 \pm 14$ & $20 \pm 14$ &$38 \pm 15$ & 34 & 21\\
$\mathrm{\Delta R\,(mas)}$ &  $64 \pm 25$ & $56 \pm 25$ &$47 \pm 26$ & 56 & 27\\
Position angle & $-45 \pm 29$ & $ 69 \pm 27$ & $55 \pm 32$ & -- & --\\
\hline

\hline
\end{tabular}
\end{center}
\end{table*}


\bibliography{g292}{}

\begin{thebibliography}{}
\expandafter\ifx\csname natexlab\endcsname\relax\def\natexlab#1{#1}\fi
\providecommand{\url}[1]{\href{#1}{#1}}
\providecommand{\dodoi}[1]{doi:~\href{http://doi.org/#1}{\nolinkurl{#1}}}
\providecommand{\doeprint}[1]{\href{http://ascl.net/#1}{\nolinkurl{http://ascl.net/#1}}}
\providecommand{\doarXiv}[1]{\href{https://arxiv.org/abs/#1}{\nolinkurl{https://arxiv.org/abs/#1}}}

\bibitem[{{Astropy Collaboration} {et~al.}(2013){Astropy Collaboration},
  {Robitaille}, {Tollerud}, {Greenfield}, {Droettboom}, {Bray}, {Aldcroft},
  {Davis}, {Ginsburg}, {Price-Whelan}, {Kerzendorf}, {Conley}, {Crighton},
  {Barbary}, {Muna}, {Ferguson}, {Grollier}, {Parikh}, {Nair}, {Unther},
  {Deil}, {Woillez}, {Conseil}, {Kramer}, {Turner}, {Singer}, {Fox}, {Weaver},
  {Zabalza}, {Edwards}, {Azalee Bostroem}, {Burke}, {Casey}, {Crawford},
  {Dencheva}, {Ely}, {Jenness}, {Labrie}, {Lim}, {Pierfederici}, {Pontzen},
  {Ptak}, {Refsdal}, {Servillat}, \& {Streicher}}]{astropy:2013}
{Astropy Collaboration}, {Robitaille}, T.~P., {Tollerud}, E.~J., {et~al.} 2013,
  \aap, 558, A33, \dodoi{10.1051/0004-6361/201322068}

\bibitem[{{Astropy Collaboration} {et~al.}(2018){Astropy Collaboration},
  {Price-Whelan}, {Sip{\H{o}}cz}, {G{\"u}nther}, {Lim}, {Crawford}, {Conseil},
  {Shupe}, {Craig}, {Dencheva}, {Ginsburg}, {Vand erPlas}, {Bradley},
  {P{\'e}rez-Su{\'a}rez}, {de Val-Borro}, {Aldcroft}, {Cruz}, {Robitaille},
  {Tollerud}, {Ardelean}, {Babej}, {Bach}, {Bachetti}, {Bakanov}, {Bamford},
  {Barentsen}, {Barmby}, {Baumbach}, {Berry}, {Biscani}, {Boquien}, {Bostroem},
  {Bouma}, {Brammer}, {Bray}, {Breytenbach}, {Buddelmeijer}, {Burke},
  {Calderone}, {Cano Rodr{\'\i}guez}, {Cara}, {Cardoso}, {Cheedella}, {Copin},
  {Corrales}, {Crichton}, {D'Avella}, {Deil}, {Depagne}, {Dietrich}, {Donath},
  {Droettboom}, {Earl}, {Erben}, {Fabbro}, {Ferreira}, {Finethy}, {Fox},
  {Garrison}, {Gibbons}, {Goldstein}, {Gommers}, {Greco}, {Greenfield},
  {Groener}, {Grollier}, {Hagen}, {Hirst}, {Homeier}, {Horton}, {Hosseinzadeh},
  {Hu}, {Hunkeler}, {Ivezi{\'c}}, {Jain}, {Jenness}, {Kanarek}, {Kendrew},
  {Kern}, {Kerzendorf}, {Khvalko}, {King}, {Kirkby}, {Kulkarni}, {Kumar},
  {Lee}, {Lenz}, {Littlefair}, {Ma}, {Macleod}, {Mastropietro}, {McCully},
  {Montagnac}, {Morris}, {Mueller}, {Mumford}, {Muna}, {Murphy}, {Nelson},
  {Nguyen}, {Ninan}, {N{\"o}the}, {Ogaz}, {Oh}, {Parejko}, {Parley}, {Pascual},
  {Patil}, {Patil}, {Plunkett}, {Prochaska}, {Rastogi}, {Reddy Janga},
  {Sabater}, {Sakurikar}, {Seifert}, {Sherbert}, {Sherwood-Taylor}, {Shih},
  {Sick}, {Silbiger}, {Singanamalla}, {Singer}, {Sladen}, {Sooley},
  {Sornarajah}, {Streicher}, {Teuben}, {Thomas}, {Tremblay}, {Turner},
  {Terr{\'o}n}, {van Kerkwijk}, {de la Vega}, {Watkins}, {Weaver}, {Whitmore},
  {Woillez}, {Zabalza}, \& {Astropy Contributors}}]{astropy:2018}
{Astropy Collaboration}, {Price-Whelan}, A.~M., {Sip{\H{o}}cz}, B.~M., {et~al.}
  2018, \aj, 156, 123, \dodoi{10.3847/1538-3881/aabc4f}

\bibitem[{{Becker} {et~al.}(2012){Becker}, {Prinz}, {Winkler}, \&
  {Petre}}]{Becker2012}
{Becker}, W., {Prinz}, T., {Winkler}, P.~F., \& {Petre}, R. 2012, \apj, 755,
  141, \dodoi{10.1088/0004-637X/755/2/141}

\bibitem[{{Bhalerao} {et~al.}(2019){Bhalerao}, {Park}, {Schenck}, {Post}, \&
  {Hughes}}]{Bhalerao2019}
{Bhalerao}, J., {Park}, S., {Schenck}, A., {Post}, S., \& {Hughes}, J.~P. 2019,
  \apj, 872, 31, \dodoi{10.3847/1538-4357/aafafd}

\bibitem[{{Burke} {et~al.}(2020){Burke}, {Laurino}, {Wmclaugh}, {Dtnguyen2},
  {Marie-Terrell}, {G{\"u}nther}, {Budynkiewicz}, {Siemiginowska}, {Aldcroft},
  {Deil}, {Sip{\H{o}}cz}, {Leinweber}, \& {Todd}}]{burke2020}
{Burke}, D., {Laurino}, O., {Wmclaugh}, {et~al.} 2020, {sherpa/sherpa: Sherpa
  4.12.1}, 4.12.1, Zenodo,  Zenodo, \dodoi{10.5281/zenodo.3944985}

\bibitem[{{Burrows} \& {Hayes}(1996)}]{Burrows1996}
{Burrows}, A., \& {Hayes}, J. 1996, \prl, 76, 352,
  \dodoi{10.1103/PhysRevLett.76.352}

\bibitem[{{Burrows} {et~al.}(2007){Burrows}, {Livne}, {Dessart}, {Ott}, \&
  {Murphy}}]{Burrows2007}
{Burrows}, A., {Livne}, E., {Dessart}, L., {Ott}, C.~D., \& {Murphy}, J. 2007,
  \apj, 655, 416, \dodoi{10.1086/509773}

\bibitem[{{Cash}(1979)}]{Cash1979}
{Cash}, W. 1979, \apj, 228, 939, \dodoi{10.1086/156922}

\bibitem[{{Chevalier}(2005)}]{Chevalier2005}
{Chevalier}, R.~A. 2005, \apj, 619, 839, \dodoi{10.1086/426584}

\bibitem[{{Clark} \& {Stephenson}(1976)}]{Clark1976}
{Clark}, D.~H., \& {Stephenson}, F.~R. 1976, \qjras, 17, 290

\bibitem[{{Davis} {et~al.}(2012){Davis}, {Bautz}, {Dewey}, {Heilmann}, {Houck},
  {Huenemoerder}, {Marshall}, {Nowak}, {Schattenburg}, {Schulz}, \&
  {Smith}}]{davis2012}
{Davis}, J.~E., {Bautz}, M.~W., {Dewey}, D., {et~al.} 2012, in Society of
  Photo-Optical Instrumentation Engineers (SPIE) Conference Series, Vol. 8443,
  Space Telescopes and Instrumentation 2012: Ultraviolet to Gamma Ray, ed.
  T.~{Takahashi}, S.~S. {Murray}, \& J.-W.~A. {den Herder}, 84431A,
  \dodoi{10.1117/12.926937}

\bibitem[{{Doe} {et~al.}(2007){Doe}, {Nguyen}, {Stawarz}, {Refsdal},
  {Siemiginowska}, {Burke}, {Evans}, {Evans}, {McDowell}, {Houck}, \&
  {Nowak}}]{doe2007}
{Doe}, S., {Nguyen}, D., {Stawarz}, C., {et~al.} 2007, in Astronomical Society
  of the Pacific Conference Series, Vol. 376, Astronomical Data Analysis
  Software and Systems XVI, ed. R.~A. {Shaw}, F.~{Hill}, \& D.~J. {Bell}, 543

\bibitem[{{Freeman} {et~al.}(2001){Freeman}, {Doe}, \&
  {Siemiginowska}}]{freeman2001}
{Freeman}, P., {Doe}, S., \& {Siemiginowska}, A. 2001, in Society of
  Photo-Optical Instrumentation Engineers (SPIE) Conference Series, Vol. 4477,
  Astronomical Data Analysis, ed. J.-L. {Starck} \& F.~D. {Murtagh}, 76--87,
  \dodoi{10.1117/12.447161}

\bibitem[{{Fruscione} {et~al.}(2006){Fruscione}, {McDowell}, {Allen},
  {Brickhouse}, {Burke}, {Davis}, {Durham}, {Elvis}, {Galle}, {Harris},
  {Huenemoerder}, {Houck}, {Ishibashi}, {Karovska}, {Nicastro}, {Noble},
  {Nowak}, {Primini}, {Siemiginowska}, {Smith}, \& {Wise}}]{fruscione2006}
{Fruscione}, A., {McDowell}, J.~C., {Allen}, G.~E., {et~al.} 2006, in Society
  of Photo-Optical Instrumentation Engineers (SPIE) Conference Series, Vol.
  6270, Society of Photo-Optical Instrumentation Engineers (SPIE) Conference
  Series, ed. D.~R. {Silva} \& R.~E. {Doxsey}, 62701V,
  \dodoi{10.1117/12.671760}

\bibitem[{{Fryer} \& {Kusenko}(2006)}]{Fryer2006}
{Fryer}, C.~L., \& {Kusenko}, A. 2006, \apjs, 163, 335, \dodoi{10.1086/500933}

\bibitem[{{Gaensler} \& {Wallace}(2003)}]{Gaensler2003}
{Gaensler}, B.~M., \& {Wallace}, B.~J. 2003, \apj, 594, 326,
  \dodoi{10.1086/376861}

\bibitem[{{Halpern} \& {Gotthelf}(2015)}]{Halpern2015}
{Halpern}, J.~P., \& {Gotthelf}, E.~V. 2015, \apj, 812, 61,
  \dodoi{10.1088/0004-637X/812/1/61}

\bibitem[{{Hobbs} {et~al.}(2005){Hobbs}, {Lorimer}, {Lyne}, \&
  {Kramer}}]{Hobbs2005}
{Hobbs}, G., {Lorimer}, D.~R., {Lyne}, A.~G., \& {Kramer}, M. 2005, \mnras,
  360, 974, \dodoi{10.1111/j.1365-2966.2005.09087.x}

\bibitem[{{Holland-Ashford} {et~al.}(2017){Holland-Ashford}, {Lopez},
  {Auchettl}, {Temim}, \& {Ramirez-Ruiz}}]{HollandAshford2017}
{Holland-Ashford}, T., {Lopez}, L.~A., {Auchettl}, K., {Temim}, T., \&
  {Ramirez-Ruiz}, E. 2017, \apj, 844, 84, \dodoi{10.3847/1538-4357/aa7a5c}

\bibitem[{{Igoshev}(2020)}]{igoshev2020I}
{Igoshev}, A.~P. 2020, \mnras, 494, 3663, \dodoi{10.1093/mnras/staa958}

\bibitem[{{Jacovich} {et~al.}(2021){Jacovich}, {Patnaude}, {Slane}, {Badenes},
  {Lee}, {Nagataki}, \& {Milisavljevic}}]{Jacovich2021}
{Jacovich}, T., {Patnaude}, D., {Slane}, P., {et~al.} 2021, \apj, 914, 41,
  \dodoi{10.3847/1538-4357/abf935}

\bibitem[{{Janka} \& {Mueller}(1994)}]{Janka1994}
{Janka}, H.~T., \& {Mueller}, E. 1994, \aap, 290, 496

\bibitem[{{Janka} {et~al.}(2022){Janka}, {Wongwathanarat}, \&
  {Kramer}}]{Janka2022}
{Janka}, H.-T., {Wongwathanarat}, A., \& {Kramer}, M. 2022, \apj, 926, 9,
  \dodoi{10.3847/1538-4357/ac403c}

\bibitem[{{Jerius} {et~al.}(2004){Jerius}, {Cohen}, {Edgar}, {Freeman},
  {Gaetz}, {Hughes}, {Nguyen}, {Podgorski}, {Tibbetts}, {Van Speybroeck}, \&
  {Zhao}}]{jerius2004}
{Jerius}, D.~H., {Cohen}, L., {Edgar}, R.~J., {et~al.} 2004, in Society of
  Photo-Optical Instrumentation Engineers (SPIE) Conference Series, Vol. 5165,
  X-Ray and Gamma-Ray Instrumentation for Astronomy XIII, ed. K.~A. {Flanagan}
  \& O.~H.~W. {Siegmund}, 402--410, \dodoi{10.1117/12.509378}

\bibitem[{{Katsuda} {et~al.}(2018){Katsuda}, {Morii}, {Janka},
  {Wongwathanarat}, {Nakamura}, {Kotake}, {Mori}, {M{\"u}ller}, {Takiwaki},
  {Tanaka}, {Tominaga}, \& {Tsunemi}}]{Katsuda2018}
{Katsuda}, S., {Morii}, M., {Janka}, H.-T., {et~al.} 2018, \apj, 856, 18,
  \dodoi{10.3847/1538-4357/aab092}

\bibitem[{{Lee} {et~al.}(2011){Lee}, {Kashyap}, {van Dyk}, {Connors}, {Drake},
  {Izem}, {Meng}, {Min}, {Park}, {Ratzlaff}, {Siemiginowska}, \&
  {Zezas}}]{Lee2011}
{Lee}, H., {Kashyap}, V.~L., {van Dyk}, D.~A., {et~al.} 2011, \apj, 731, 126,
  \dodoi{10.1088/0004-637X/731/2/126}

\bibitem[{{Lee} {et~al.}(2010){Lee}, {Park}, {Hughes}, {Slane}, {Gaensler},
  {Ghavamian}, \& {Burrows}}]{Lee2010}
{Lee}, J.-J., {Park}, S., {Hughes}, J.~P., {et~al.} 2010, \apj, 711, 861,
  \dodoi{10.1088/0004-637X/711/2/861}

\bibitem[{{Mayer} {et~al.}(2020){Mayer}, {Becker}, {Patnaude}, {Winkler}, \&
  {Kraft}}]{Mayer2020}
{Mayer}, M., {Becker}, W., {Patnaude}, D., {Winkler}, P.~F., \& {Kraft}, R.
  2020, \apj, 899, 138, \dodoi{10.3847/1538-4357/aba121}

\bibitem[{{Mayer} \& {Becker}(2021)}]{Mayer2021}
{Mayer}, M. G.~F., \& {Becker}, W. 2021, \aap, 651, A40,
  \dodoi{10.1051/0004-6361/202141119}

\bibitem[{{Mignani} {et~al.}(2007){Mignani}, {de Luca}, {Zaggia}, {Sester},
  {Pellizzoni}, {Mereghetti}, \& {Caraveo}}]{Mignani2007}
{Mignani}, R.~P., {de Luca}, A., {Zaggia}, S., {et~al.} 2007, \aap, 473, 883,
  \dodoi{10.1051/0004-6361:20077768}

\bibitem[{{Mignani} {et~al.}(2019){Mignani}, {De Luca}, {Zharikov}, {Hummel},
  {Becker}, \& {Pellizzoni}}]{Mignani2019}
{Mignani}, R.~P., {De Luca}, A., {Zharikov}, S., {et~al.} 2019, \mnras, 486,
  5716, \dodoi{10.1093/mnras/stz1195}

\bibitem[{{Nakamura} {et~al.}(2019){Nakamura}, {Takiwaki}, \&
  {Kotake}}]{Nakamura2019}
{Nakamura}, K., {Takiwaki}, T., \& {Kotake}, K. 2019, \pasj, 71, 98,
  \dodoi{10.1093/pasj/psz080}

\bibitem[{{Ng} \& {Romani}(2007)}]{ng2007}
{Ng}, C.~Y., \& {Romani}, R.~W. 2007, \apj, 660, 1357, \dodoi{10.1086/513597}

\bibitem[{{Noutsos} {et~al.}(2012){Noutsos}, {Kramer}, {Carr}, \&
  {Johnston}}]{noutsos2012}
{Noutsos}, A., {Kramer}, M., {Carr}, P., \& {Johnston}, S. 2012, \mnras, 423,
  2736, \dodoi{10.1111/j.1365-2966.2012.21083.x}

\bibitem[{{Park} {et~al.}(2007){Park}, {Hughes}, {Slane}, {Burrows},
  {Gaensler}, \& {Ghavamian}}]{Park2007}
{Park}, S., {Hughes}, J.~P., {Slane}, P.~O., {et~al.} 2007, \apjl, 670, L121,
  \dodoi{10.1086/524406}

\bibitem[{{Socrates} {et~al.}(2005){Socrates}, {Blaes}, {Hungerford}, \&
  {Fryer}}]{Socrates2005}
{Socrates}, A., {Blaes}, O., {Hungerford}, A., \& {Fryer}, C.~L. 2005, \apj,
  632, 531, \dodoi{10.1086/431786}

\bibitem[{{Stephenson} \& {Green}(2002)}]{Stephenson2002}
{Stephenson}, F.~R., \& {Green}, D.~A. 2002, Historical supernovae and their
  remnants, 5

\bibitem[{{Temim} {et~al.}(2017){Temim}, {Slane}, {Plucinsky}, {Gelfand},
  {Castro}, \& {Kolb}}]{Temim2017}
{Temim}, T., {Slane}, P., {Plucinsky}, P.~P., {et~al.} 2017, \apj, 851, 128,
  \dodoi{10.3847/1538-4357/aa9d41}

\bibitem[{{Winkler} {et~al.}(2009){Winkler}, {Twelker}, {Reith}, \&
  {Long}}]{Winkler2009}
{Winkler}, P.~F., {Twelker}, K., {Reith}, C.~N., \& {Long}, K.~S. 2009, \apj,
  692, 1489, \dodoi{10.1088/0004-637X/692/2/1489}

\bibitem[{{Wongwathanarat} {et~al.}(2010){Wongwathanarat}, {Janka}, \&
  {M{\"u}ller}}]{Wongwathanarat2010}
{Wongwathanarat}, A., {Janka}, H.-T., \& {M{\"u}ller}, E. 2010, \apjl, 725,
  L106, \dodoi{10.1088/2041-8205/725/1/L106}

\bibitem[{{Wongwathanarat} {et~al.}(2013){Wongwathanarat}, {Janka}, \&
  {M{\"u}ller}}]{Wongwathanarat2013}
{Wongwathanarat}, A., {Janka}, H.~T., \& {M{\"u}ller}, E. 2013, \aap, 552,
  A126, \dodoi{10.1051/0004-6361/201220636}

\bibitem[{{Woosley}(1987)}]{Woosley1987}
{Woosley}, S.~E. 1987, in The Origin and Evolution of Neutron Stars, ed. D.~J.
  {Helfand} \& J.~H. {Huang}, Vol. 125, 255

\bibitem[{{Yao} {et~al.}(2021){Yao}, {Zhu}, {Manchester}, {Coles}, {Li},
  {Wang}, {Kramer}, {Stinebring}, {Feng}, {Yan}, {Miao}, {Yuan}, {Wang}, \&
  {Lu}}]{yao2021}
{Yao}, J., {Zhu}, W., {Manchester}, R.~N., {et~al.} 2021, Nature Astronomy, 5,
  788, \dodoi{10.1038/s41550-021-01360-w}

\end{thebibliography}
\bibliographystyle{aasjournal}


\end{CJK*}
\end{document}